\documentclass[superscriptaddress,twocolumn]{revtex4-1}
\usepackage{amsmath,amssymb,graphicx}
\usepackage{hyperref}
\usepackage[all]{hypcap}
\hypersetup{
  colorlinks   = true, 
  urlcolor     = blue, 
  linkcolor    = blue, 
  citecolor   = blue, 
}

\begin{document}

\title{Nagaoka ferromagnetism observed in a quantum dot plaquette}

\author{J. P. Dehollain}
\thanks{These authors contributed equally to this work.}
\affiliation{QuTech and Kavli Institute of Nanoscience, TU Delft, P.O. Box 5046, 2600 GA Delft, The Netherlands.}
\affiliation{Present address: School of Mathematical and Physical Sciences, University of Technology Sydney, NSW 2026, Australia.}
\author{U. Mukhopadhyay}
\thanks{These authors contributed equally to this work.}
\affiliation{QuTech and Kavli Institute of Nanoscience, TU Delft, P.O. Box 5046, 2600 GA Delft, The Netherlands.}
\author{V. P. Michal}
\affiliation{QuTech and Kavli Institute of Nanoscience, TU Delft, P.O. Box 5046, 2600 GA Delft, The Netherlands.}
\author{Y. Wang}
\author{B. Wunsch}
\affiliation{Department of Physics, Harvard University, Cambridge 02138, USA.}
\author{C. Reichl}
\author{W. Wegscheider}
\affiliation{Solid State Physics Laboratory, ETH Z\"{u}rich, Z\"{u}rich 8093, Switzerland.}
\author{M. S. Rudner}
\affiliation{Center for Quantum Devices and Niels Bohr International Academy, Niels Bohr Institute, University of Copenhagen, 2100 Copenhagen, Denmark.}
\author{E. Demler}
\affiliation{Department of Physics, Harvard University, Cambridge 02138, USA.}
\author{L. M. K. Vandersypen}
\email[e-mail: ]{l.m.k.vandersypen@tudelft.nl}
\affiliation{QuTech and Kavli Institute of Nanoscience, TU Delft, P.O. Box 5046, 2600 GA Delft, The Netherlands.}

\date{\today}

\maketitle

{\bfseries
    Engineered, highly-controllable quantum systems hold promise as simulators of emergent physics beyond the capabilities of classical computers~\cite{Feynman1982}. An important problem in many-body physics is \textit{itinerant magnetism}, which originates purely from long-range interactions of free electrons and whose existence in real systems has been subject to debate for decades~\cite{Auerbach1994,Mattis2006}. Here we use a quantum simulator consisting of a four-site square plaquette of quantum dots~\cite{Mukhopadhyay2018} to demonstrate Nagaoka ferromagnetism~\cite{Nagaoka1966}. This form of itinerant magnetism has been rigorously studied theoretically~\cite{Mattis2003,Nielsen2007,Oguri2007,Stecher2010} but has remained unattainable in experiment. We load the plaquette with three electrons and demonstrate the predicted emergence of spontaneous ferromagnetic correlations through pairwise measurements of spin. We find the ferromagnetic ground state is remarkably robust to engineered disorder in the on-site potentials and can induce a transition to the low-spin state by changing the plaquette topology to an open chain. This demonstration of Nagaoka ferromagnetism highlights that quantum simulators can be used to study physical phenomena that have not yet been observed in any system before. The work also constitutes an important step towards large-scale quantum dot simulators of correlated electron systems.
}

The potential impact of discovering and understanding exotic forms of magnetism and superconductivity is one of the largest motivations for research in condensed-matter physics. These quantum mechanically governed effects result from the strong correlations that arise between interacting electrons. Modelling and simulating such systems can in some instances only be achieved through the use of engineered, controllable systems that operate in the quantum regime~\cite{Feynman1982}. Efforts to build quantum simulators have already demonstrated great promise at this early stage~\cite{Georgescu2014}, mainly led by the ultracold atom community~\cite{Trotzky2008,Nascimbene2012,Bloch2012,Dai2017,Goerg2018,Salomon2018,Nichols2018}. More broadly, quantum simulations of many-body fermionic systems have been carried out in a range of experimental systems such as quantum dot lattices~\cite{Singha2011}, dopant atoms~\cite{Salfi2016}, superconducting circuits~\cite{Barends2015} and trapped ions~\cite{Martinez2016}. 

Electrostatically defined semiconductor quantum dots~\cite{Kouwenhoven2001,Wiel2002,Hanson2007} have been proposed as excellent candidates for quantum simulations~\cite{Manousakis2002,Byrnes2008,Barthelemy2013}. Their ability to reach thermal energies far below the hopping and on-site interaction energies enable access to previously unexplored material phases. Quantum dot systems have already achieved success in realising simulations of Mott-insulator physics in linear arrays~\cite{Hensgens2017}. Additionally, the feasibility to extend these systems into 2D lattices has recently been demonstrated~\cite{Thalineau2012,Seo2013,Noiri2017,Mukhopadhyay2018,Mortemousque2018}, including the ability to perform measurements of spin correlations~\cite{Mukhopadhyay2018}. As a result, quantum dot systems are now prime candidates for exploring how superconductivity and magnetism emerge in strongly-correlated electron systems~\cite{Scalapino1995,Tsuei2000,Balents2010}.

The emergence of magnetism in purely itinerant electron systems presents a long-standing problem in quantum many-body physics~\cite{Auerbach1994,Mattis2006} with only few rigorous theoretical results, for instance in systems with special flat bands or Nagaoka's ferromagnetism (see Ref.~\cite{Tasaki1998} and references therein). The Nagaoka model of ferromagnetism~\cite{Thouless1965,Nagaoka1966} relies on the simplicity of the Hubbard model~\cite{Hubbard1963}, which captures complex correlations between electrons in a lattice with only two Hamiltonian parameters. Using this single-band model, Nagaoka proved analytically that for some lattice configurations, and in the limit of infinitely strong interactions, the presence of a single hole on top of a Mott-insulating state with one electron per site renders the ground state ferromagnetic. The Nagaoka mechanism can be intuitively understood as an interference effect between the different paths that the hole can take across the lattice. These paths interfere constructively when all lattice sites have the same spin orientation, which lowers the kinetic energy of the hole.

\begin{figure}
    \includegraphics[width=\columnwidth]{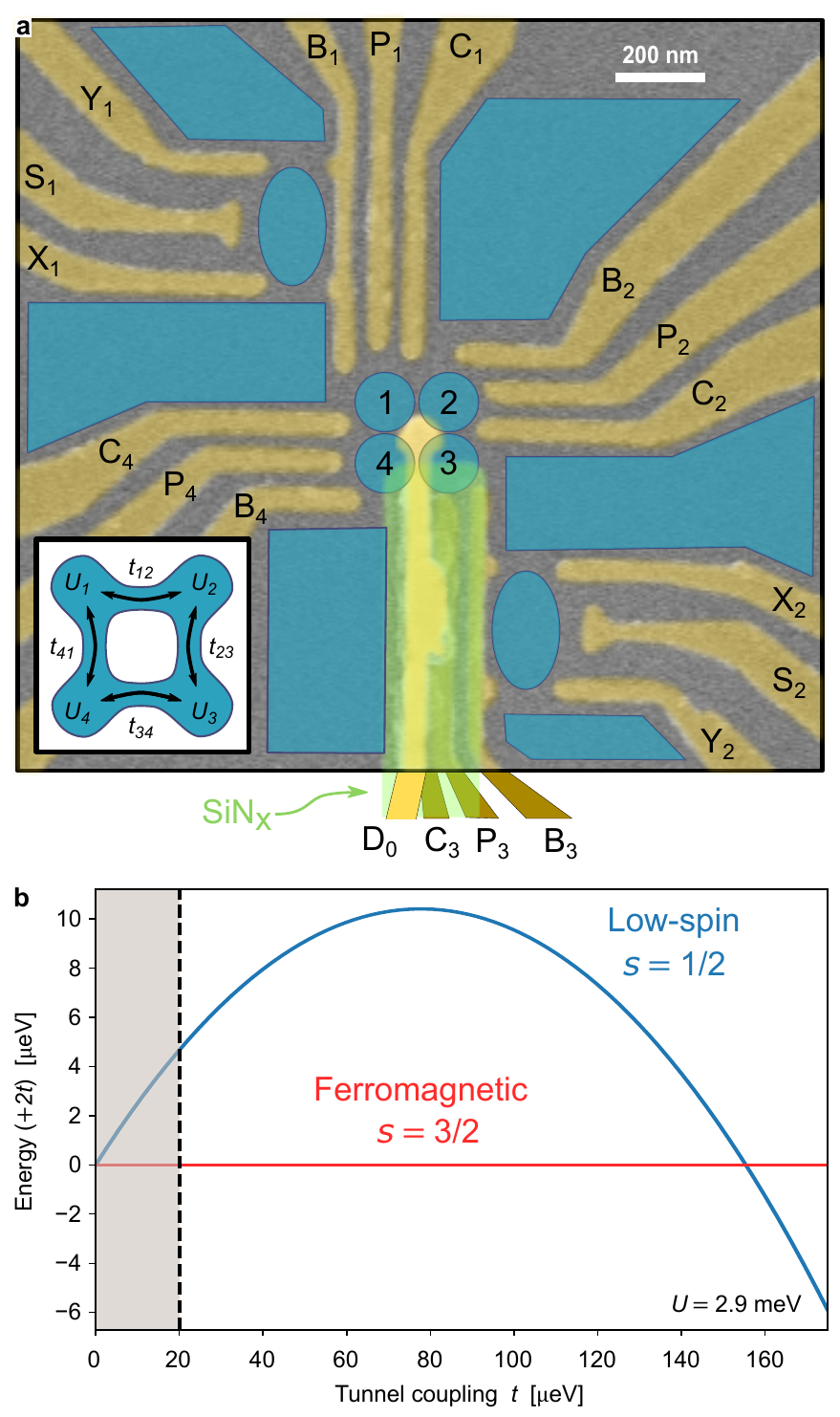}
    \caption{\textbf{Device schematic and Nagaoka model.} \textbf{a,} False coloured scanning electron microscope image of a device from the same batch as the one used in the experiments. The gate structure used to define the quantum dots is coloured in dark gold. A slab of silicon nitride (coloured in green) is laid over gates $C_3$ and $P_3$, to electrically isolate those gates from the $D_0$ gate (coloured in bright gold) which runs over them and contacts the substrate at the centre of the structure. A sketch of the expected 2DEG density in blue shows the 4 dots forming a plaquette in the centre of the device, along with nearby charge sensors and electron reservoirs. \textbf{b,} Energy spectrum as a function of tunnel coupling using the solution expressed in Eq.~\ref{eq:energies}, with $U = 2.9$~meV. Shaded area shows the experimentally accessible range of $t$ in this system. \label{fig:model}}
\end{figure}

Given that Nagaoka obtained his rigorous result using unrealistic limits, it has been an open question whether this mechanism can still be responsible for the observation of ferromagnetism in an experimental, finite-size system, in the presence of long-range interactions and disorder, as well as additional available orbitals. In this light, we note  that a ferromagnetic state is a fully-polarised spin state, and as such is an eigenstate of the total spin operator $S_{tot}^2$. This statement is true whether the system is in the thermodynamic limit, or whether it is finite size. The feasibility of performing a quantum simulation of Nagaoka ferromagnetism has been explored theoretically for quantum dots~\cite{Mattis2003,Nielsen2007,Oguri2007} as well as optical superlattices~\cite{Stecher2010}, but there are no experimental reports to date.

In this article, we present clear experimental evidence of Nagaoka ferromagnetism, using a quantum dot device designed to host a 2$\times$2 array of electrons. Using the high degree of parameter tunability, we study how external magnetic fields and disorder in local potentials affect the magnetic nature of the ground state. Furthermore, by effectively tuning the geometry of the system from periodic to open boundary conditions, we experimentally demonstrate the suppression of ferromagnetism expected from the Lieb-Mattis theorem~\cite{Lieb1962}.

\section*{Nagaoka in the quantum dot plaquette}
    The single-band Hubbard model provides a simple description of interacting electrons in a lattice, such as the plaquette of electrostatically defined and controlled quantum dots (Fig.~\ref{fig:model}a). The Hamiltonian contains competing kinetic energy and electron-electron interaction terms:
    \begin{equation}
        \mathcal{H}_{H} = - \sum_{\langle i,j\rangle \sigma} t_{i,j} c^\dagger_{i\sigma} c_{j\sigma} + \sum_i U_i n_{i\uparrow} n_{i\downarrow} - \sum_i \mu_i n_i, \label{eq:hub}
    \end{equation}
    where $t_{i,j}$ describes electron tunnelling between sites $i$ and $j$, $U_i$ is the on-site Coulomb repulsion energy at site $i$ and $\mu_i$ is a local energy offset at site $i$. In typical quantum dot systems, $U_i$ is mainly set by the geometry of the device and is on the order of $1$~meV, while $t_{i,j}$ and $\mu_i$ can be controlled by gate voltages in the range of $0$ to $0.5$~meV and $0$ to $20$~meV respectively~\cite{Barthelemy2013}. The operators $c_{i\sigma}$, $c_{i\sigma}^{\dagger}$ and $n_{i\sigma}$ represent the second quantisation annihilation, creation and number operators for an electron on site $i$ with spin projection $\sigma = \{ \uparrow,\downarrow \}$.
    
    Nagaoka ferromagnetism is predicted to occur with an almost-half-filled lattice, which for the case of the 2$\times$2 plaquette corresponds to having three interacting electrons in the four-site system. By additionally restricting the system to nearest-neighbour-only coupling, the Hamiltonian is analytically solvable~\cite{Mattis2003} for homogeneous interactions ($U_i = U$, $t_{i,j} = t$, $\mu_i = 0$) and in the limit $U \gg t$, where the lowest eigenenergies are:
    \begin{equation}
    E_{3/2} = -2t \quad \mathrm{and} \quad E_{1/2} = -\sqrt{3} t - \frac{5t^2}{U}. \label{eq:energies}
    \end{equation}
    Here, $E_{3/2}$ is the energy of the high-spin, ferromagnetic quadruplets (with total spin $s = 3/2$) and $E_{1/2}$ is the energy of the 2 sets of low-spin $s = 1/2$ degenerate doublets (see supplementary material for details).
    
\begin{figure*}
    \includegraphics[width=\textwidth]{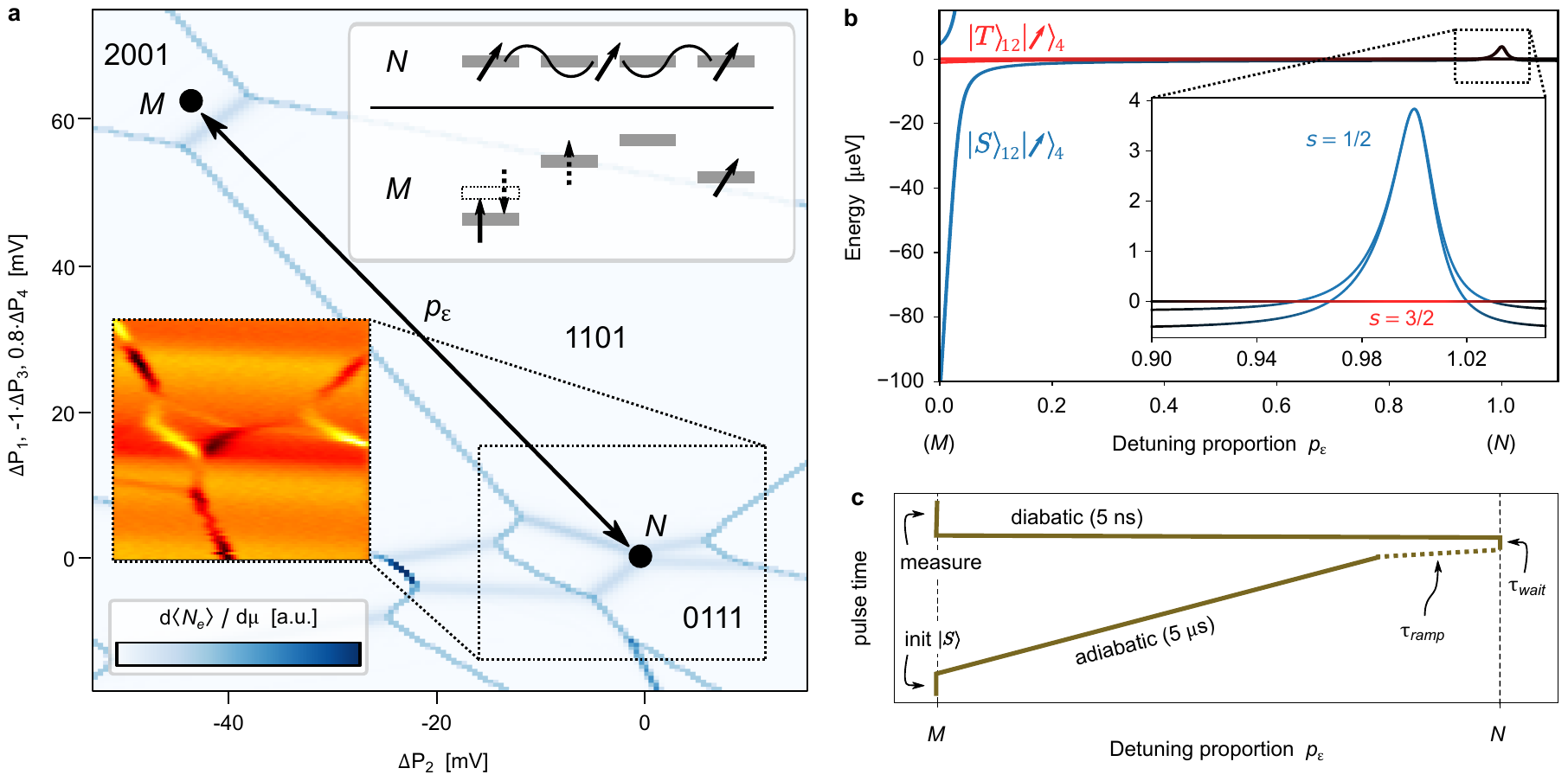}
    \caption{\textbf{Experimental protocol.} \textbf{a,} Simulated charge stability diagram plotting the change in electron occupation in the approximate gate space used in the experiment. In the experiment we pulse in a straight line in gate space from point~$M$ to point~$N$ and back. Top-right inset shows a schematic of the local energies at points~$N$ and $M$, highlighting in the latter how the measurement of 2 spins in the singlet-triplet basis is performed through spin-to-charge conversion. Lower-left inset shows a measured charge stability diagram of the dotted region, with the same gate voltage ratios as the simulation, which we use in the experiment to calibrate the gate voltages at point~$N$. \textbf{b,} Calculated energy spectrum as a function of detuning proportion, using the theoretical model (Eq.~\ref{eq:hub} and supplementary text) without spin-coupling effects. Parameters were set to the experimental values presented in the methods. Inset shows a zoomed-in spectrum of the region where the 3 spins are delocalised on all 4 dots, where there are a total of 8 states: the $s = 3/2$ quadruplets (red) and the 2 sets of $s = 1/2$ doublets (blue), of which one set connects with the $\lvert T \rangle$ branch and the other with the $\lvert S \rangle$ branch at point~$M$. Line colours represent the spin state of the system in each region, denoted by the labels in the figure. The energies extracted from the numerical solutions are offset by the energy of $\lvert {s,m} \rangle = \lvert {3/2,+3/2} \rangle$. \textbf{c,} Pulse sequence used in the experiment (see methods for detailed description). \label{fig:method}}
\end{figure*}

    We note that the Hamiltonian in Eq.~\ref{eq:hub} neglects some of the essential features of the experimental device used in this work. For comparison with experimental results, we employ a more general model Hamiltonian, in which we account for interdot Coulomb repulsion (in Fig.~\ref{fig:method}a), spin-orbit and hyperfine interactions (in Fig.~S4a), as well as the effects of external magnetic fields (in Fig.~\ref{fig:field}a-b). The implementation of these terms is described in detail in the supplementary material. In addition to this model, we have also performed an ab initio calculation (see supplementary material and Ref.~\cite{Wang2019}) based on multiple orbitals solved from a potential landscape with 2$\times$2 minima, showing very similar results to those obtained with Eq.~\ref{eq:hub}.

    The simple model described by Eqs.~\ref{eq:hub}~and~\ref{eq:energies} already provides some useful insight into the parameter regimes relevant to the experiment. The ferromagnetic state is the ground state at large $U/t$, with a transition to a low-spin ground state occurring at $U/t = 18.7$. The quantum dot array used in this work has an average $U \approx 2.9$~meV, with tunable nearest-neighbour tunnel couplings in the range of $0 < t \lesssim 20$~$\mu$eV~\cite{Mukhopadhyay2018}. Unless otherwise stated, we set $t_{i,i+1} \approx 16$~$\mu$eV. This means that we are probing the regime where the ground state is expected to be ferromagnetic (see Fig.~\ref{fig:model}b).
    
    We prepare the system by using charge stability diagrams~\cite{Wiel2006} to find the appropriate voltage bias that will stabilise the system in a charge configuration with 3 resonant electrons delocalised in the 4 sites. We set the local energy reference at this regime as $\mu_i(N) = 0$~eV for all dots, and refer to this condition as point~$N$. Charge stability diagrams are also used to tune the gates to the measurement point~$M$, where single-shot measurements in the singlet-triplet basis are performed on 2 of the 3 electrons. Fig.~\ref{fig:method}a and Fig.~S1 show simulated and measured charge stability diagrams where points $N$ and $M$ can be identified, along with an inset schematic of the dot local energies at these points.
    
\begin{figure}
    \includegraphics[width=\columnwidth]{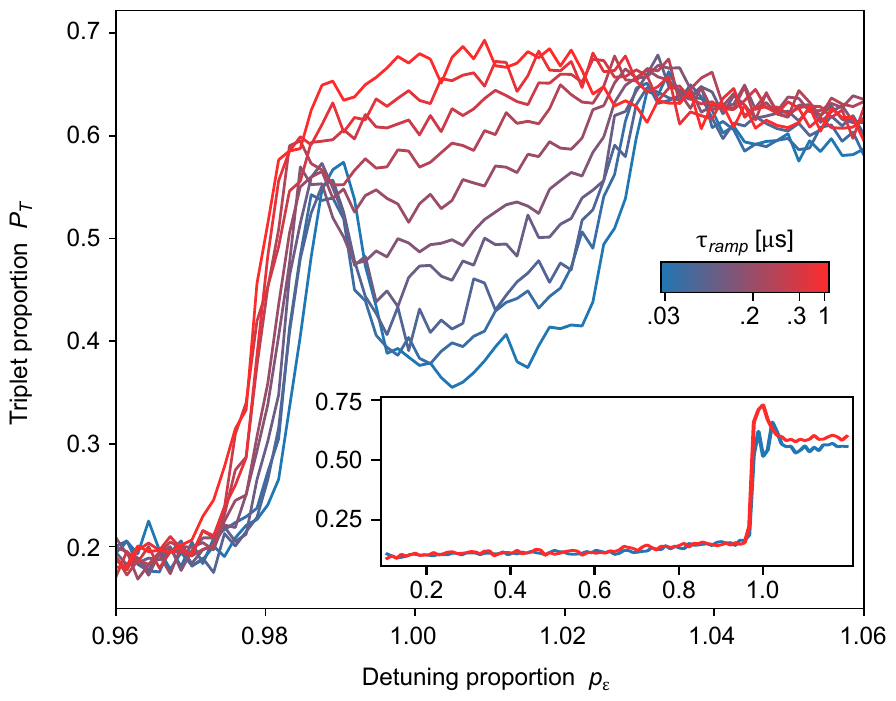}
    \caption{\textbf{Main signature of Nagaoka ferromagnetism.} Measured $P_T$ vs $p_\epsilon$ using the protocol described in the main text (10000 repetitions per point in each curve). Different curves correspond to different values of $\tau_{ramp}$. The main figure focuses on the region close to point~$N$, while the inset is zoomed out to the entire detuning range for the 2 extreme values of $\tau_{ramp}$. $\tau_{wait}$ is fixed to $50$~ns ($500$~ns) for the main figure (inset). \label{fig:main}}
\end{figure}

    With the accessible system parameters, the theoretically expected (Fig.~\ref{fig:model}b) energy gap between the ferromagnetic ($s = 3/2$) and low-spin ($s = 1/2$) states at point~$N$ is $E_{1/2} - E_{3/2} \approx 4$~$\mu$eV, comparable to the measured electron temperature $k_B T_e \approx 6$~$\mu$eV~($70$~mK)~\cite{Mukhopadhyay2018}. In order to study the magnetic properties of the ground state, we have developed a technique (see Fig.~\ref{fig:method}b-c) based on initialising a low-entropy state at point~$M$ and adiabatically pulsing to point~$N$ to access the ground state. We then diabatically pulse back to point~$M$ and probe the spin state of the system on timescales faster than the relaxation times. By shortening the ramp time $\tau_{ramp}$ of the pulse from point~$M$ to $N$ we can also access excited states. To distinguish whether the system is in a ferromagnetic or low-spin state, we repeat the cycle of preparation and measurement, and extract the triplet probability $P_T$, which informs us on the nature of the original 3-spin state (see supplementary material for details). In the Methods section we provide a detailed description of these preparation and measurement protocols.

\begin{figure}
    \includegraphics[width=\columnwidth]{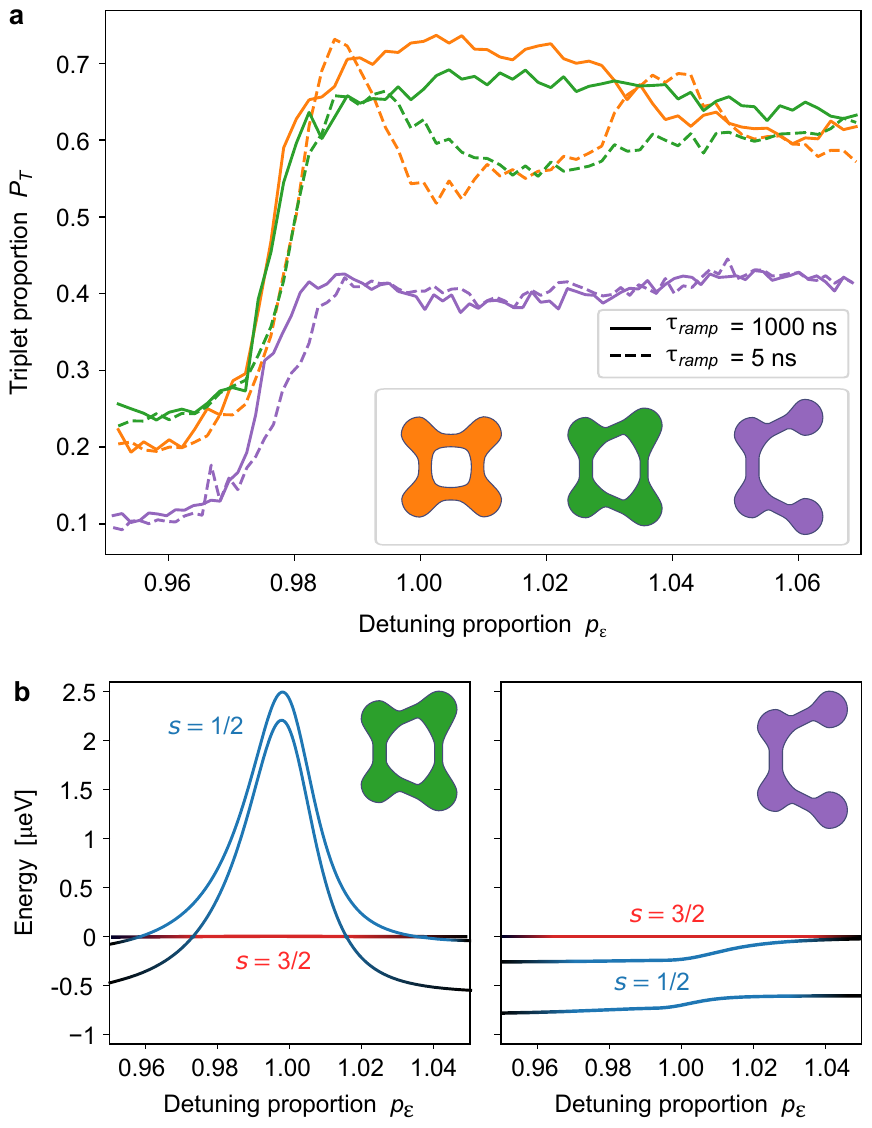}
	\caption{\textbf{From ring to chain.} \textbf{a,} Comparison of 3 measurements with the following values of tunnel couplings $[t_{1,2}, t_{2,3}, t_{3,4}, t_{4,1}]$ (in $\mu$eV): $[19(1), 15(1), 17(2), 19(5)]$~(orange); $[16(1), 7.9(5), 20(2), 19(2)]$~(green); $[18(4), 0.0, 21(1), 21(3)]$~(purple). The offsets in $P_T$ between the curves are not attributed to the topology, but are due to small measurement-to-measurement variations in the thermal excitation rate during the initialisation stage of the protocol. \textbf{b,} Calculated energy spectrum as a function of detuning proportion, using the tunnel coupling values corresponding to the green (left) and purple (right) plots from \textbf{a}. \label{fig:chain}}
\end{figure}

\section*{Experimental results}
    Fig.~\ref{fig:main} shows plots of $P_T$ as we perform measurements at different values of detuning proportion $p_\epsilon$, defined as the quantity that sets the linear combination of gate voltages $P_i$ (see Fig.~\ref{fig:model}a) such that at $p_\epsilon = 1 (0)$ the system is tuned to point~$N$($M$). From the inset of the figure we highlight that $P_T$ remains at a low value for most of the range, with a sharp increase as $p_\epsilon$ approaches 1 (point~$N$). This is consistent with the expectation that the electrons will remain localised until the region close to point~$N$, where they begin to delocalise and the $P_T$ measurement starts to project the three interacting spins (see Methods for details). This is expected to happen after $p_\epsilon \approx 0.96$, where the energy spectrum (see inset of Fig.~\ref{fig:method}b) shows an energy level crossing and the $s = 3/2$ states become the ground state. The non-zero triplet fraction at low $p_\epsilon$ is attributed partly to thermal excitations during the initialisation stage (see Methods)--as a consequence of the finite electron temperature--and partly to a small probability of leakage to excited states during the pulse.
	
	The main plot shows the measurement around point~$N$, for a range of $\tau_{ramp}$. In the region $0.99 < p_\epsilon < 1.03$, a clear increase of $P_T$ is observed as $\tau_{ramp}$ is increased, consistent with a gradual transition from diabatically pulsing into the low-spin state, to adiabatically pulsing into the ferromagnetic state, where $P_T$ is maximum. For the faster pulses, we see `peaks' of $P_T$ at $p_\epsilon = 0.99$ and $1.03$, where the pulse reaches the energy-level crossings, as all the spin states quickly (i.e., much faster than the experimental timescales) mix due to the nuclear hyperfine fields and spin-orbit coupling~\cite{Khaetskii2002,Merkulov2002,Chekhovich2013}.
	
\begin{figure}
	\includegraphics[width=\columnwidth]{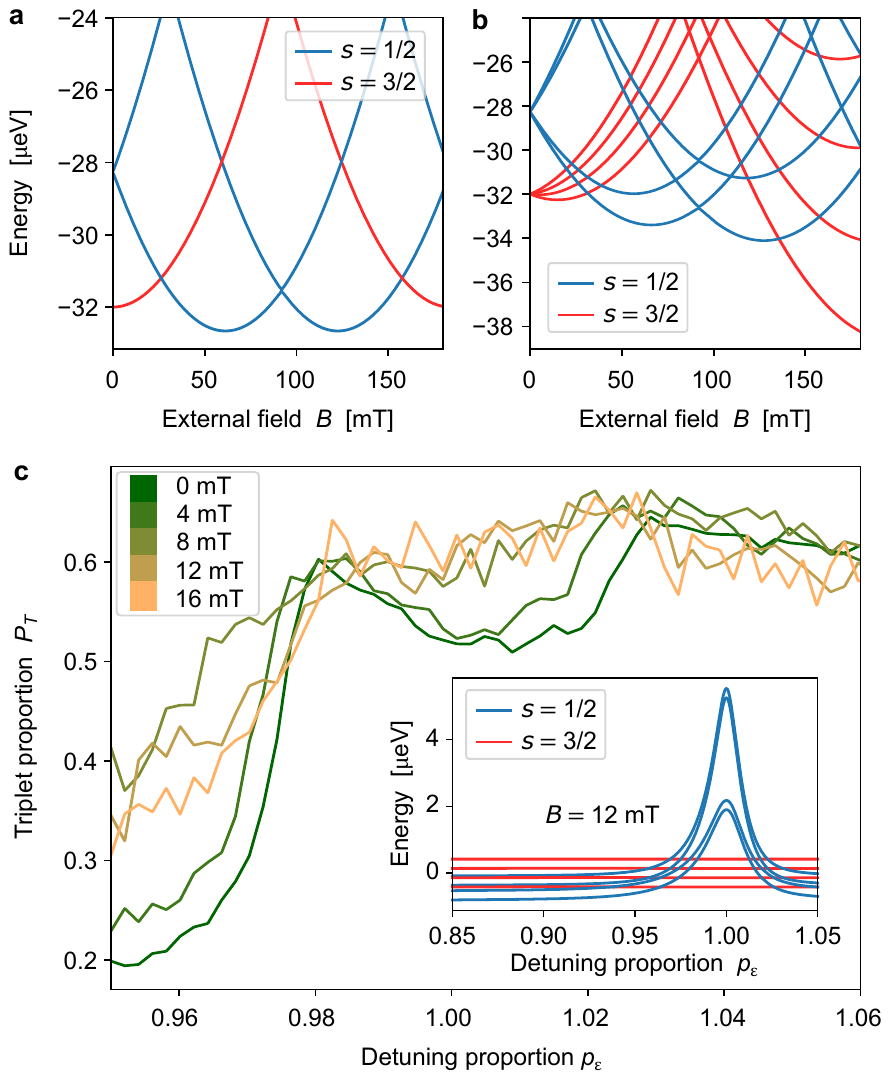}
	\caption{\textbf{Applying an external magnetic field.} \textbf{a,} Lowest eigenenergies of the $s = 1/2$ (blue) and $s = 3/2$ (red) states at point~$N$ as a function of magnetic field, obtained from the numerical model after including the effect of an Aharonov-Bohm phase. \textbf{b,} Same as \textbf{a} but with the addition of the Zeeman effect, and the lowest 4 eigenenergies of each $s$ states are shown. \textbf{c,} Experimental measurement using diabatic passage, for different fields in the range of $0$ to $16$~mT. Inset shows a numerically calculated spectrum at $12$~mT, with the Aharonov-Bohm phase and Zeeman effect included in the model. \label{fig:field}}
\end{figure}

	The $\tau_{ramp}$ timescale for the diabatic to adiabatic transition shown in Fig.~S3a can be theoretically studied using time-evolution simulations with an extended-Hubbard model (see supplementary material for details). From fits to the data we estimate a hyperfine coupling parameter $\delta_N = 73 \pm 3$~neV, in agreement with previous observations and calculations in similar GaAs quantum dot systems~\cite{Petta2005,Koppens2005,Chekhovich2013}. Fig.~S3b shows $P_T$ as a function of the waiting time $\tau_{wait}$ spent at point~$N$ ($p_\epsilon = 1$), consistent with thermal equilibration of the system with a timescale $\tau_{relax} \sim 2$~$\mu$s.
	
    We note that we cannot directly assign the measured values of $P_T$ to $s = 1/2$ and $s = 3/2$ populations, because the observed $P_T$ is subject to measurement imperfections caused by mechanisms that are difficult to disentangle, such as the finite measurement bandwidth, the signal-to-noise ratio and $\lvert T \rangle$ to $\lvert S \rangle$ relaxation, as well as unwanted leakage to other states during the pulsed passages.

\subsection*{Changing topology -- from 2D to 1D } 
	Whereas the square plaquette can be thought of as a 1D ring, the \emph{Lieb-Mattis theorem}~\cite{Lieb1962,Mattis2003} asserts that the ground state of a 1D array of electrons with open boundary conditions has the lowest possible spin. We can intuitively understand the difference between these two configurations when we consider how the hole tunnels to its next-nearest neighbour~\cite{Mattis2006}. In a 2D plaquette, the hole has 2 possible paths to the next-nearest neighbour. If the system is initialised in any of the $s = 1/2$ configurations, the 2 paths will leave the system in 2 different spin configurations. On the other hand, for an $s = 3/2$ system the 2 paths leave identical spin configurations, and interfere constructively to lower the kinetic energy. In contrast, in an open boundary 1D array, the kinetic energy of the hole is independent of the spin configurations of the neighbouring electrons as there is only one path for the hole through the array.

	One powerful feature of the quantum dot system is that the tunnel barriers can be tuned independently, allowing us to test different array topologies. In Fig.~\ref{fig:chain}a we compare diabatic and adiabatic sweeps as we raise the tunnel barrier that controls $t_{23}$, effectively transforming the plaquette into a system that behaves more like an open-boundary 1D system. In the latter regime, we see that $P_T$ becomes insensitive to sweep rate. Additionally, we no longer observe the peaks of $P_T$ for the fast sweep rate, which we had associated with mixing at the avoided level crossings. From these observations we infer that for the open chain, the instantaneous ground state does not exhibit an avoided crossing between an $s = 1/2$ state and an $s = 3/2$ state as the system is taken to point~$N$. In this regime the $p_\epsilon$ sweeps will always evolve to the $s = 1/2$ ground state, independent of the sweep rate. This interpretation is also consistent with the numerical simulations of the energy spectrum shown in Fig.~\ref{fig:chain}b.

\subsection*{Effects of external magnetic fields } 
	Given that Nagaoka ferromagnetism originates from interference effects due to the trajectories of the hole around the ring, it then follows that a magnetic flux through the plaquette will add an Aharonov-Bohm phase~\cite{Aharonov1959} that disturbs the interference effects. We capture this effect in the theoretical model by adding a magnetic field dependent gauge to the tunneling term in Eq.~\ref{eq:hub}. In addition, the application of an external field subjects the system to the Zeeman effect, causing a spin-dependent energy offset. See supplementary material for details on how the gauge and Zeeman terms are implemented in the extended-Hubbard model.
	 
\begin{figure*}
	\includegraphics[width=\textwidth]{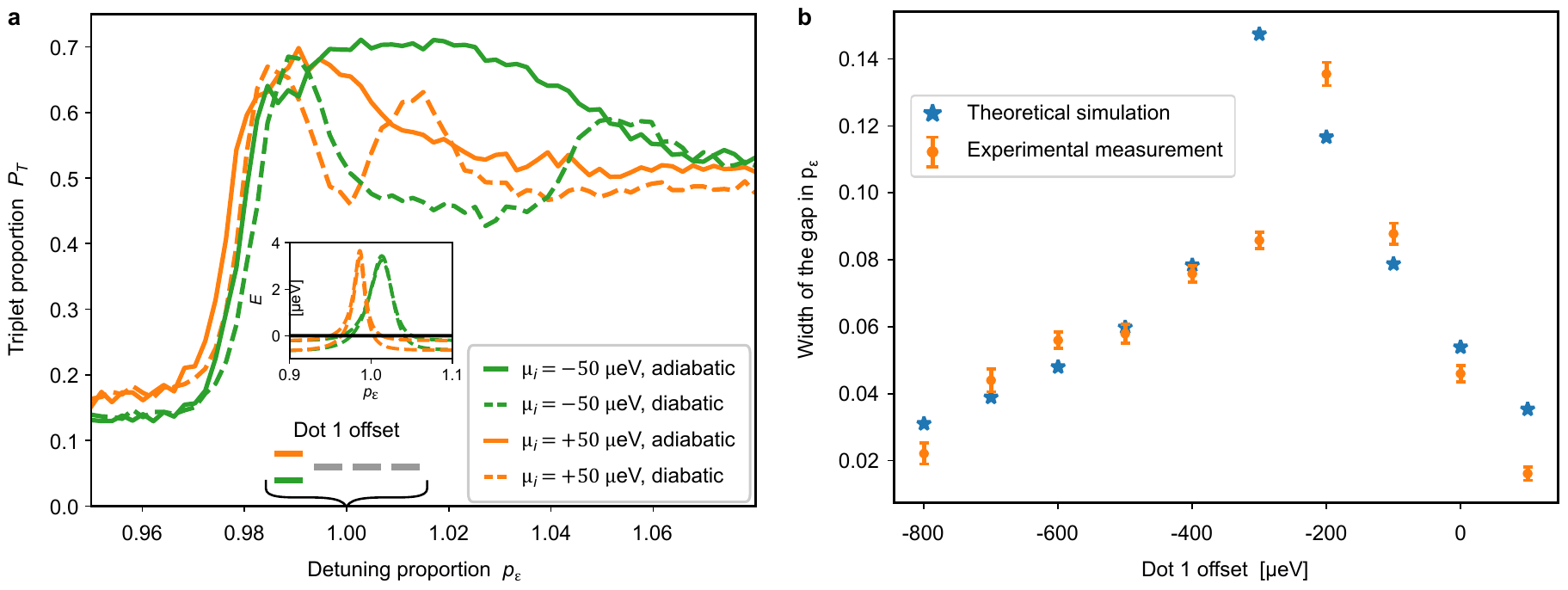}
	\caption{\textbf{Local energy offsets.} \textbf{a,} Experimental measurements with a similar protocol used to probe the states at point~$N$, but pulsing instead to a point~$N'$, where the local energy of dot 1 is offset by $\pm 50$~$\mu$eV. Inset shows numerically calculated spectra for the same experimental condition. \textbf{b,} Comparison between the experimentally measured and theoretically simulated ranges in detuning proportion where the $s = 3/2$ to $s = 1/2$ gap is visible. The points used to determine the widths in both experimental measurements and theoretical simulations are shown in Fig.~S8. Error bars are calculated from the fits to the peaks used to determine the widths from the experimental data.}\label{fig:offsets}
\end{figure*}

	Fig.~\ref{fig:field}a shows the effect of a magnetic field through the plaquette on the spectrum, ignoring the Zeeman effect. The lowest $s = 1/2$ and $s = 3/2$ levels at point~$N$ are shown as a function of the applied field, where periodic crossings can be observed. In the range $30 < B < 160$~mT, the system ground state transitions to the low-spin state, with the perhaps counterintuitive implication that we can destroy the ferromagnetic state by applying a magnetic field. Additionally, this effect highlights that the ferromagnetic state in this system is dominated by the Nagaoka effect rather than long-range interactions. In line with this observation, the ab initio calculations suggest that long-range interactions only account for $\sim 20\%$ of the ferromagnetic polarisation~\cite{Wang2019}. When we include the Zeeman effect (see Fig.~\ref{fig:field}b) the picture becomes more complicated, because both Zeeman and orbital effects cause perturbations of similar energy scales.

	From this initial numerical analysis, it is clear that the experimental characterisation of the effect of the external field will be challenging, due to the increased complexity of the spectral structure of the spin states as a function of field. The small energy splittings that appear both at point~$N$, as well as at lower $p_\epsilon$ values (see inset of Fig.~\ref{fig:field}c) are expected to cause mixing of the spin states during the adiabatic pulses. To minimise this mixing, we adjusted the pulsing protocol such that we pulse adiabatically ($1$~$\mu$s ramp) to $p_\epsilon = 0.2$, then pulse diabatically ($5$~ns ramp) the rest of the way. The results in Fig.~\ref{fig:field}c show that from $4$ to $8$~mT $P_T$ increases at point~$N$, and we stop observing the characteristic dip. Note that the range of field that we were able to probe is still below the estimated ground state transition point ($\sim 30$~mT). Therefore, we infer that the observed increase in $P_T$ results from hybridisation of the $s = 1/2$ and $s = 3/2$ states as their energy gap reduces. We cannot claim that the observed hybridisation of states is occurring solely at point~$N$, as it is evident from the increase in $P_T$ at $p_\epsilon < 0.97$ (i.e.~prior to the energy-level crossings) that some of the mixing is occurring during the pulse. However, we do see that $P_T$ in all plots converge at the energy-level crossings ($p_\epsilon \approx 0.97$ and $p_\epsilon \approx 1.03$) suggesting that the Aharonov-Bohm orbital effects are partly responsible for the additional mixing near point~$N$. Attempts to perform the measurement at higher fields closer to the expected spin-state transition resulted in similar plots.
\subsection*{Sensitivity to local energy offsets } 
	We also use the tunability available in quantum dot systems to study the effects of disorder of the local potential. For the plot in Fig.~\ref{fig:offsets}a, we modified the experimental protocol used to probe the states at point~$N$, pulsing instead to a point $N'$, where the local energy of dot 1 is offset by $\pm 50$~$\mu$eV. We achieve this by employing the \emph{virtual gates} technique~\cite{Hensgens2017,Mukhopadhyay2018}, which gives access to control knobs that map a linear combination of $P_i$ gates onto local dot energy offsets. The measurements show that the region in the detuning trajectory where the ferromagnetic state is the ground state changes in width and position when different offsets are applied. The panel inset shows the expected energy spectra when we simulate the experimental conditions using the model in Eq.~\ref{eq:hub}. The spectra show excellent qualitative agreement to the measured variations in width and position of the gapped region. In Fig.~S7 we repeated the measurement on each of the four dots, showing similar qualitative agreement with theory.
	
	Fig.~\ref{fig:offsets}b compares experimental measurements and theoretical predictions of the width of the detuning proportion region (as defined in the caption of Fig.~S8) with a ferromagnetic ground state energy gap, for offsets of dot 1 in the range $+100$ to $-800$~$\mu$eV. This plot further confirms the interpretation of the experimental observations, showing excellent agreement between measurements and theoretical predictions. Remarkably, the system still shows signs of the ferromagnetic ground state with offsets up to $-400$~$\mu$eV (see Fig.~S8), more than an order of magnitude larger than the tunnel coupling. We also note that the theoretical simulations in Fig.~S8 suggest that from $-500$~$\mu$eV the ground state is no longer $s = 3/2$, even though the measurement still shows a gap in $P_T$ between diabatic and adiabatic sweeps. The presence of this gap is explained by the large splitting between the two $s = 1/2$ branches that occurs at such large local offsets.
\section*{Discussion}
    In this work we have presented the first measurements showing experimental evidence of Nagaoka's 50-year old theory in a small-scale system. The large degree of tunability, high ratio of interaction strength to temperature, and fast measurement techniques available to quantum dot systems, allowed observing both the ferromagnetic ground state and the low-spin excited state of an almost-half-filled lattice of electrons. By performing a quantum simulation involving both charge and spin, it builds on previous demonstrations~\cite{Hensgens2017} that quantum dot systems can be useful simulators of the extended Hubbard model, despite their initial inhomogeneities in the potential shape and local energies. Furthermore, in this work we showed a flavour of the capabilities for studying the sensitivity to disorder, and these experiments already revealed some remarkable effects, when we found that the Nagaoka condition can still be observed after offsetting a local energy by amounts much larger than the tunnel coupling. This can readily be studied in further detail, along with other possibilities for exploring the effects of disorder, which could bring insights into e.g.,~the stability of the ferromagnetic state. More quantitative insight of the energy gap between the spin states can be achieved through spectroscopy measurements, using techniques such as applying oscillating electric fields through a gate~\cite{Nowack2007} or observing `exchange-like' oscillations~\cite{Malinowski2018}.
    
    While the problem of three electrons in a four-site plaquette can be solved analytically using the single-band Hubbard picture, a complete description of this experimental system that includes all its available orbitals is not easily tractable, analytically or numerically. Indeed, the computational cost of the ab initio calculation of the 5 lowest states, with long-range and on-site interaction terms being considered, is on the order of $10000$ CPU hours. Small-scale simulations on tractable models can be used to systematically benchmark the performance of devices as the scale-up technology develops towards devices that can perform classically intractable simulations. Larger quantum dot systems (or other experimentally controllable systems), such as 2$\times$N or M$\times$N arrays can shed more light on the existence of purely itinerant ferromagnetism in real systems. The exchange interaction grows proportionally to the system size, creating a competition against the hopping energy that is characteristic of Nagaoka ferromagnetism, and leaving the fate of the Nagaoka mechanism in larger systems unknown.
    

\begin{thebibliography}{10}
\expandafter\ifx\csname url\endcsname\relax
  \def\url#1{\texttt{#1}}\fi
\expandafter\ifx\csname urlprefix\endcsname\relax\def\urlprefix{URL }\fi
\providecommand{\bibinfo}[2]{#2}
\providecommand{\eprint}[2][]{\url{#2}}

\bibitem{Feynman1982}
\bibinfo{author}{Feynman, R.~P.}
\newblock \bibinfo{title}{{S}imulating physics with computers}.
\newblock \emph{\bibinfo{journal}{Int. J. Theor. Phys}}
  \textbf{\bibinfo{volume}{21}}, \bibinfo{pages}{467--488}
  (\bibinfo{year}{1982}).

\bibitem{Auerbach1994}
\bibinfo{author}{Auerbach, A.}
\newblock \emph{\bibinfo{title}{{I}nteracting {E}lectrons and {Q}uantum
  {M}agnetism}} (\bibinfo{publisher}{{S}pringer {N}ew {Y}ork},
  \bibinfo{year}{1994}).

\bibitem{Mattis2006}
\bibinfo{author}{Mattis, D.~C.}
\newblock \emph{\bibinfo{title}{{T}he {T}heory of {M}agnetism {M}ade {S}imple}}
  (\bibinfo{publisher}{{W}orld {S}cientific}, \bibinfo{year}{2006}).

\bibitem{Mukhopadhyay2018}
\bibinfo{author}{Mukhopadhyay, U.}, \bibinfo{author}{Dehollain, J.~P.},
  \bibinfo{author}{Reichl, C.}, \bibinfo{author}{Wegscheider, W.} \&
  \bibinfo{author}{Vandersypen, L. M.~K.}
\newblock \bibinfo{title}{A 2x2 quantum dot array with controllable inter-dot
  tunnel couplings}.
\newblock \emph{\bibinfo{journal}{Appl. Phys. Lett.}}
  \textbf{\bibinfo{volume}{112}}, \bibinfo{pages}{183505}
  (\bibinfo{year}{2018}).

\bibitem{Nagaoka1966}
\bibinfo{author}{Nagaoka, Y.}
\newblock \bibinfo{title}{Ferromagnetism in a narrow, almost half-filled s
  band}.
\newblock \emph{\bibinfo{journal}{Phys. Rev.}} \textbf{\bibinfo{volume}{147}},
  \bibinfo{pages}{392--405} (\bibinfo{year}{1966}).

\bibitem{Mattis2003}
\bibinfo{author}{Mattis, D.~C.}
\newblock \bibinfo{title}{Eigenvalues and magnetism of electrons on an
  artificial molecule}.
\newblock \emph{\bibinfo{journal}{Int. J. Nanosci.}}
  \textbf{\bibinfo{volume}{02}}, \bibinfo{pages}{165--170}
  (\bibinfo{year}{2003}).

\bibitem{Nielsen2007}
\bibinfo{author}{Nielsen, E.} \& \bibinfo{author}{Bhatt, R.~N.}
\newblock \bibinfo{title}{Nanoscale ferromagnetism in nonmagnetic doped
  semiconductors}.
\newblock \emph{\bibinfo{journal}{Phys. Rev. B}} \textbf{\bibinfo{volume}{76}},
  \bibinfo{pages}{161202(R)} (\bibinfo{year}{2007}).

\bibitem{Oguri2007}
\bibinfo{author}{Oguri, A.}, \bibinfo{author}{Nisikawa, Y.},
  \bibinfo{author}{Tanaka, Y.} \& \bibinfo{author}{Numata, T.}
\newblock \bibinfo{title}{Kondo screening of a high-spin {N}agaoka state in a
  triangular quantum dot}.
\newblock \emph{\bibinfo{journal}{J. Magn. Magn. Mater.}}
  \textbf{\bibinfo{volume}{310}}, \bibinfo{pages}{1139--1141}
  (\bibinfo{year}{2007}).

\bibitem{Stecher2010}
\bibinfo{author}{von Stecher, J.}, \bibinfo{author}{Demler, E.},
  \bibinfo{author}{Lukin, M.~D.} \& \bibinfo{author}{Rey, A.~M.}
\newblock \bibinfo{title}{Probing interaction-induced ferromagnetism in optical
  superlattices}.
\newblock \emph{\bibinfo{journal}{New J. Phys.}} \textbf{\bibinfo{volume}{12}},
  \bibinfo{pages}{055009} (\bibinfo{year}{2010}).

\bibitem{Georgescu2014}
\bibinfo{author}{Georgescu, I.~M.}, \bibinfo{author}{Ashhab, S.} \&
  \bibinfo{author}{Nori, F.}
\newblock \bibinfo{title}{Quantum simulation}.
\newblock \emph{\bibinfo{journal}{Rev. Mod. Phys.}}
  \textbf{\bibinfo{volume}{86}}, \bibinfo{pages}{153--185}
  (\bibinfo{year}{2014}).

\bibitem{Trotzky2008}
\bibinfo{author}{Trotzky, S.} \emph{et~al.}
\newblock \bibinfo{title}{Time-resolved observation and control of
  superexchange interactions with ultracold atoms in optical lattices}.
\newblock \emph{\bibinfo{journal}{Science}} \textbf{\bibinfo{volume}{319}},
  \bibinfo{pages}{295--299} (\bibinfo{year}{2008}).

\bibitem{Nascimbene2012}
\bibinfo{author}{Nascimb{\`{e}}ne, S.} \emph{et~al.}
\newblock \bibinfo{title}{Experimental realization of plaquette resonating
  valence-bond states with ultracold atoms in optical superlattices}.
\newblock \emph{\bibinfo{journal}{Phys. Rev. Lett.}}
  \textbf{\bibinfo{volume}{108}}, \bibinfo{pages}{205301}
  (\bibinfo{year}{2012}).

\bibitem{Bloch2012}
\bibinfo{author}{Bloch, I.}, \bibinfo{author}{Dalibard, J.} \&
  \bibinfo{author}{Nascimb{\`{e}}ne, S.}
\newblock \bibinfo{title}{Quantum simulations with ultracold quantum gases}.
\newblock \emph{\bibinfo{journal}{Nat. Phys.}} \textbf{\bibinfo{volume}{8}},
  \bibinfo{pages}{267--276} (\bibinfo{year}{2012}).

\bibitem{Dai2017}
\bibinfo{author}{Dai, H.-N.} \emph{et~al.}
\newblock \bibinfo{title}{Four-body ring-exchange interactions and anyonic
  statistics within a minimal toric-code {H}amiltonian}.
\newblock \emph{\bibinfo{journal}{Nat. Phys.}} \textbf{\bibinfo{volume}{13}},
  \bibinfo{pages}{1195--1200} (\bibinfo{year}{2017}).

\bibitem{Goerg2018}
\bibinfo{author}{G\"{o}rg, F.} \emph{et~al.}
\newblock \bibinfo{title}{Enhancement and sign change of magnetic correlations
  in a driven quantum many-body system}.
\newblock \emph{\bibinfo{journal}{Nature}} \textbf{\bibinfo{volume}{553}},
  \bibinfo{pages}{481--485} (\bibinfo{year}{2018}).

\bibitem{Salomon2018}
\bibinfo{author}{Salomon, G.} \emph{et~al.}
\newblock \bibinfo{title}{Direct observation of incommensurate magnetism in
  {H}ubbard chains}.
\newblock \emph{\bibinfo{journal}{Nature}} \textbf{\bibinfo{volume}{565}},
  \bibinfo{pages}{56--60} (\bibinfo{year}{2018}).

\bibitem{Nichols2018}
\bibinfo{author}{Nichols, M.~A.} \emph{et~al.}
\newblock \bibinfo{title}{Spin transport in a {M}ott insulator of ultracold
  fermions}.
\newblock \emph{\bibinfo{journal}{Science}} \textbf{\bibinfo{volume}{363}},
  \bibinfo{pages}{383--387} (\bibinfo{year}{2018}).

\bibitem{Singha2011}
\bibinfo{author}{Singha, A.} \emph{et~al.}
\newblock \bibinfo{title}{Two-dimensional {M}ott-{H}ubbard electrons in an
  artificial honeycomb lattice}.
\newblock \emph{\bibinfo{journal}{Science}} \textbf{\bibinfo{volume}{332}},
  \bibinfo{pages}{1176--1179} (\bibinfo{year}{2011}).

\bibitem{Salfi2016}
\bibinfo{author}{Salfi, J.} \emph{et~al.}
\newblock \bibinfo{title}{Quantum simulation of the {H}ubbard model with dopant
  atoms in silicon}.
\newblock \emph{\bibinfo{journal}{Nat. Commun.}} \textbf{\bibinfo{volume}{7}},
  \bibinfo{pages}{11342} (\bibinfo{year}{2016}).

\bibitem{Barends2015}
\bibinfo{author}{Barends, R.} \emph{et~al.}
\newblock \bibinfo{title}{Digital quantum simulation of fermionic models with a
  superconducting circuit}.
\newblock \emph{\bibinfo{journal}{Nat. Commun.}} \textbf{\bibinfo{volume}{6}},
  \bibinfo{pages}{7654} (\bibinfo{year}{2015}).

\bibitem{Martinez2016}
\bibinfo{author}{Martinez, E.~A.} \emph{et~al.}
\newblock \bibinfo{title}{Real-time dynamics of lattice gauge theories with a
  few-qubit quantum computer}.
\newblock \emph{\bibinfo{journal}{Nature}} \textbf{\bibinfo{volume}{534}},
  \bibinfo{pages}{516--519} (\bibinfo{year}{2016}).

\bibitem{Kouwenhoven2001}
\bibinfo{author}{Kouwenhoven, L.~P.}, \bibinfo{author}{Austing, D.~G.} \&
  \bibinfo{author}{Tarucha, S.}
\newblock \bibinfo{title}{Few-electron quantum dots}.
\newblock \emph{\bibinfo{journal}{Rep. Prog. Phys.}}
  \textbf{\bibinfo{volume}{64}}, \bibinfo{pages}{701--736}
  (\bibinfo{year}{2001}).

\bibitem{Wiel2002}
\bibinfo{author}{van~der Wiel, W.~G.} \emph{et~al.}
\newblock \bibinfo{title}{Electron transport through double quantum dots}.
\newblock \emph{\bibinfo{journal}{Rev. Mod. Phys.}}
  \textbf{\bibinfo{volume}{75}}, \bibinfo{pages}{1--22} (\bibinfo{year}{2002}).

\bibitem{Hanson2007}
\bibinfo{author}{Hanson, R.}, \bibinfo{author}{Kouwenhoven, L.~P.},
  \bibinfo{author}{Petta, J.~R.}, \bibinfo{author}{Tarucha, S.} \&
  \bibinfo{author}{Vandersypen, L. M.~K.}
\newblock \bibinfo{title}{{S}pins in few-electron quantum dots}.
\newblock \emph{\bibinfo{journal}{Rev. Mod. Phys.}}
  \textbf{\bibinfo{volume}{79}}, \bibinfo{pages}{1217--1265}
  (\bibinfo{year}{2007}).

\bibitem{Manousakis2002}
\bibinfo{author}{Manousakis, E.}
\newblock \bibinfo{title}{A quantum-dot array as model for copper-oxide
  superconductors: A dedicated quantum simulator for the many-fermion problem}.
\newblock \emph{\bibinfo{journal}{J. Low Temp. Phys.}}
  \textbf{\bibinfo{volume}{126}}, \bibinfo{pages}{1501--1513}
  (\bibinfo{year}{2002}).

\bibitem{Byrnes2008}
\bibinfo{author}{Byrnes, T.}, \bibinfo{author}{Kim, N.~Y.},
  \bibinfo{author}{Kusudo, K.} \& \bibinfo{author}{Yamamoto, Y.}
\newblock \bibinfo{title}{Quantum simulation of {F}ermi-{H}ubbard models in
  semiconductor quantum-dot arrays}.
\newblock \emph{\bibinfo{journal}{Phys. Rev. B}} \textbf{\bibinfo{volume}{78}},
  \bibinfo{pages}{075320} (\bibinfo{year}{2008}).

\bibitem{Barthelemy2013}
\bibinfo{author}{Barthelemy, P.} \& \bibinfo{author}{Vandersypen, L. M.~K.}
\newblock \bibinfo{title}{Quantum dot systems: a versatile platform for quantum
  simulations}.
\newblock \emph{\bibinfo{journal}{Ann. Phys. (Leipzig)}}
  \textbf{\bibinfo{volume}{525}}, \bibinfo{pages}{808--826}
  (\bibinfo{year}{2013}).

\bibitem{Hensgens2017}
\bibinfo{author}{Hensgens, T.} \emph{et~al.}
\newblock \bibinfo{title}{Quantum simulation of a {F}ermi{\textendash}{H}ubbard
  model using a semiconductor quantum dot array}.
\newblock \emph{\bibinfo{journal}{Nature}} \textbf{\bibinfo{volume}{548}},
  \bibinfo{pages}{70--73} (\bibinfo{year}{2017}).
\newblock \eprint{1702.07511v1}.

\bibitem{Thalineau2012}
\bibinfo{author}{Thalineau, R.} \emph{et~al.}
\newblock \bibinfo{title}{A few-electron quadruple quantum dot in a closed
  loop}.
\newblock \emph{\bibinfo{journal}{Appl. Phys. Lett.}}
  \textbf{\bibinfo{volume}{101}}, \bibinfo{pages}{103102}
  (\bibinfo{year}{2012}).

\bibitem{Seo2013}
\bibinfo{author}{Seo, M.} \emph{et~al.}
\newblock \bibinfo{title}{Charge frustration in a triangular triple quantum
  dot}.
\newblock \emph{\bibinfo{journal}{Phys. Rev. Lett.}}
  \textbf{\bibinfo{volume}{110}} (\bibinfo{year}{2013}).

\bibitem{Noiri2017}
\bibinfo{author}{Noiri, A.} \emph{et~al.}
\newblock \bibinfo{title}{A triangular triple quantum dot with tunable tunnel
  couplings}.
\newblock \emph{\bibinfo{journal}{Semicond. Sci. Technol.}}
  \textbf{\bibinfo{volume}{32}}, \bibinfo{pages}{084004}
  (\bibinfo{year}{2017}).

\bibitem{Mortemousque2018}
\bibinfo{author}{Mortemousque, P.-A.} \emph{et~al.}
\newblock \bibinfo{title}{Coherent control of individual electron spins in a
  two dimensional array of quantum dots}.
\newblock \emph{\bibinfo{journal}{http://arxiv.org/abs/1808.06180v1}}
  (\bibinfo{year}{2018}).
\newblock \eprint{http://arxiv.org/abs/1808.06180v1}.

\bibitem{Scalapino1995}
\bibinfo{author}{Scalapino, D.~J.}
\newblock \bibinfo{title}{The case for dx2 - y2 pairing in the cuprate
  superconductors}.
\newblock \emph{\bibinfo{journal}{Phys. Rev.}} \textbf{\bibinfo{volume}{250}},
  \bibinfo{pages}{329--365} (\bibinfo{year}{1995}).

\bibitem{Tsuei2000}
\bibinfo{author}{Tsuei, C.~C.} \& \bibinfo{author}{Kirtley, J.~R.}
\newblock \bibinfo{title}{Pairing symmetry in cuprate superconductors}.
\newblock \emph{\bibinfo{journal}{Rev. Mod. Phys.}}
  \textbf{\bibinfo{volume}{72}}, \bibinfo{pages}{969--1016}
  (\bibinfo{year}{2000}).

\bibitem{Balents2010}
\bibinfo{author}{Balents, L.}
\newblock \bibinfo{title}{Spin liquids in frustrated magnets}.
\newblock \emph{\bibinfo{journal}{Nature}} \textbf{\bibinfo{volume}{464}},
  \bibinfo{pages}{199--208} (\bibinfo{year}{2010}).

\bibitem{Tasaki1998}
\bibinfo{author}{Tasaki, H.}
\newblock \bibinfo{title}{From {N}agaoka's ferromagnetism to flat-band
  ferromagnetism and beyond: {A}n introduction to ferromagnetism in the
  {H}ubbard model}.
\newblock \emph{\bibinfo{journal}{Progress of Theoretical Physics}}
  \textbf{\bibinfo{volume}{99}}, \bibinfo{pages}{489--548}
  (\bibinfo{year}{1998}).

\bibitem{Thouless1965}
\bibinfo{author}{Thouless, D.~J.}
\newblock \bibinfo{title}{Exchange in solid {3He} and the {H}eisenberg
  {H}amiltonian}.
\newblock \emph{\bibinfo{journal}{Planet. Space Sci.}}
  \textbf{\bibinfo{volume}{86}}, \bibinfo{pages}{893--904}
  (\bibinfo{year}{1965}).

\bibitem{Hubbard1963}
\bibinfo{author}{Hubbard, J.}
\newblock \bibinfo{title}{Electron correlations in narrow energy bands}.
\newblock \emph{\bibinfo{journal}{P. Roy. Soc. Lon. A}}
  \textbf{\bibinfo{volume}{276}}, \bibinfo{pages}{238--257}
  (\bibinfo{year}{1963}).

\bibitem{Lieb1962}
\bibinfo{author}{Lieb, E.} \& \bibinfo{author}{Mattis, D.}
\newblock \bibinfo{title}{Theory of ferromagnetism and the ordering of
  electronic energy levels}.
\newblock \emph{\bibinfo{journal}{Phys. Rev.}} \textbf{\bibinfo{volume}{125}},
  \bibinfo{pages}{164--172} (\bibinfo{year}{1962}).

\bibitem{Wang2019}
\bibinfo{author}{Wang, Y.} \emph{et~al.}
\newblock \bibinfo{title}{Ab initio exact diagonalization simulation of the
  {N}agaoka transition in quantum dots}.
\newblock \emph{\bibinfo{journal}{Phys. Rev. B}}
  \textbf{\bibinfo{volume}{100}}, \bibinfo{pages}{155133}
  (\bibinfo{year}{2019}).

\bibitem{Wiel2006}
\bibinfo{author}{van~der Wiel, W.~G.}, \bibinfo{author}{Stopa, M.},
  \bibinfo{author}{Kodera, T.}, \bibinfo{author}{Hatano, T.} \&
  \bibinfo{author}{Tarucha, S.}
\newblock \bibinfo{title}{{S}emiconductor quantum dots for electron spin
  qubits}.
\newblock \emph{\bibinfo{journal}{New J. Phys.}} \textbf{\bibinfo{volume}{8}},
  \bibinfo{pages}{28} (\bibinfo{year}{2006}).

\bibitem{Khaetskii2002}
\bibinfo{author}{Khaetskii, A.~V.}, \bibinfo{author}{Loss, D.} \&
  \bibinfo{author}{Glazman, L.}
\newblock \bibinfo{title}{Electron spin decoherence in quantum dots due to
  interaction with nuclei}.
\newblock \emph{\bibinfo{journal}{Phys. Rev. Lett.}}
  \textbf{\bibinfo{volume}{88}}, \bibinfo{pages}{186802}
  (\bibinfo{year}{2002}).

\bibitem{Merkulov2002}
\bibinfo{author}{Merkulov, I.~A.}, \bibinfo{author}{Efros, A.~L.} \&
  \bibinfo{author}{Rosen, M.}
\newblock \bibinfo{title}{Electron spin relaxation by nuclei in semiconductor
  quantum dots}.
\newblock \emph{\bibinfo{journal}{Phys. Rev. B}} \textbf{\bibinfo{volume}{65}},
  \bibinfo{pages}{205309} (\bibinfo{year}{2002}).

\bibitem{Chekhovich2013}
\bibinfo{author}{Chekhovich, E.~A.} \emph{et~al.}
\newblock \bibinfo{title}{Nuclear spin effects in semiconductor quantum dots}.
\newblock \emph{\bibinfo{journal}{Nat. Mater.}} \textbf{\bibinfo{volume}{12}},
  \bibinfo{pages}{494--504} (\bibinfo{year}{2013}).

\bibitem{Petta2005}
\bibinfo{author}{Petta, J.~R.}
\newblock \bibinfo{title}{{C}oherent {M}anipulation of {C}oupled {E}lectron
  {S}pins in {S}emiconductor {Q}uantum {D}ots}.
\newblock \emph{\bibinfo{journal}{Science}} \textbf{\bibinfo{volume}{309}},
  \bibinfo{pages}{2180--2184} (\bibinfo{year}{2005}).

\bibitem{Koppens2005}
\bibinfo{author}{Koppens, F. H.~L.}
\newblock \bibinfo{title}{{C}ontrol and {D}etection of {S}inglet-{T}riplet
  {M}ixing in a {R}andom {N}uclear {F}ield}.
\newblock \emph{\bibinfo{journal}{Science}} \textbf{\bibinfo{volume}{309}},
  \bibinfo{pages}{1346--1350} (\bibinfo{year}{2005}).

\bibitem{Aharonov1959}
\bibinfo{author}{Aharonov, Y.} \& \bibinfo{author}{Bohm, D.}
\newblock \bibinfo{title}{Significance of electromagnetic potentials in the
  quantum theory}.
\newblock \emph{\bibinfo{journal}{Phys. Rev.}} \textbf{\bibinfo{volume}{115}},
  \bibinfo{pages}{485--491} (\bibinfo{year}{1959}).

\bibitem{Nowack2007}
\bibinfo{author}{Nowack, K.~C.}, \bibinfo{author}{Koppens, F. H.~L.},
  \bibinfo{author}{Nazarov, Y.~V.} \& \bibinfo{author}{Vandersypen, L. M.~K.}
\newblock \bibinfo{title}{{C}oherent {C}ontrol of a {S}ingle {E}lectron {S}pin
  with {E}lectric {F}ields}.
\newblock \emph{\bibinfo{journal}{Science}} \textbf{\bibinfo{volume}{318}},
  \bibinfo{pages}{1430--1433} (\bibinfo{year}{2007}).

\bibitem{Malinowski2018}
\bibinfo{author}{Malinowski, F.~K.} \emph{et~al.}
\newblock \bibinfo{title}{Spin of a multielectron quantum dot and its interaction with a neighboring electron}.
\newblock \emph{\bibinfo{journal}{Phys. Rev. X}} \textbf{\bibinfo{volume}{8}},
\bibinfo{pages}{011045} (\bibinfo{year}{2018}).

\end{thebibliography}

\vspace{-10pt}
\section*{Acknowledgements}
We acknowledge input and discussions with M. Chan, S. Philips, Y. Nazarov, F. Liu, L. Janssen, T. Hensgens, T. Fujita and all of the Vandersypen team, as well as experimental support by L. Blom, C. van Diepen, P. Eendebak, R. Schouten, R. Vermeulen, R. van Ooijik, H. van der Does, M. Ammerlaan, J. Haanstra, S. Visser and R. Roeleveld. L.M.K.V. thanks the NSF-funded MIT-Harvard Center for Ultracold Atoms for its hospitality. This work was supported by grants from the Netherlands Organisation for Scientific Research (FOM projectruimte and NWO Vici) (J.P.D., U.M., L.M.K.V.), the European Research Council (ERC-Synergy) (V.P.M., L.M.K.V.), the Postdoctoral Fellowship in Quantum Science of the Harvard-MPQ Center for Quantum Optics and AFOSR-MURI Quantum Phases of Matter (Grant No.~FA9550-14-1-0035) (Y.W.), the Swiss National Science Foundation (C.R., W.W.), The Villum Foundation (M.S.R.).\\

\textbf{Data and code availability.} Datasets obtained from the measurements described in this work, as well as code to plot the datasets and implement models used to reproduce all the figure in the main manuscript are available in the repository Zenodo with the identifier 10.5281/zenodo.3258940.

\end{document}


\title{Supplementary material:\\Nagaoka ferromagnetism observed in a quantum dot plaquette}

\author{Juan P. Dehollain}
\author{Uditendu Mukhopadhyay}
\author{Vincent P. Michal}
\author{Yao Wang}
\author{Bernhard Wunsch}
\author{Christian Reichl}
\author{Werner Wegscheider}
\author{Mark S. Rudner}
\author{Eugene Demler}
\author{Lieven M.K. Vandersypen}

\maketitle

\section*{Experimental methods}
    \subsection*{Device fabrication}
    The experiment was performed using an array of four gate-defined quantum dots in a 2$\times$2 geometry~\cite{Mukhopadhyay2018}. The device substrate consists of an AlGaAs/GaAs heterostructure, designed to have a 2-dimensional electron gas (2DEG) $90$~nm below the surface. The quantum dots are defined and controlled using metallic gates patterned on the surface of the substrate, as shown in the scanning electron microscope image of a device from the same batch as the one used in this work in Fig.~1a. We employed a double-layer gate structure to form this dot array. The first layer of gates--which includes all gates except $D_0$--was created using electron-beam lithography, evaporation and lift-off of Ti/Au with 5/20~nm thickness. We then fabricate a 1.5x0.2~$\mu$m dielectric slab on top of the gates $C_3$ and $P_3$, using electron-beam lithography, sputtering and lift-off of SiN${_x}$ with 50~nm thickness. Finally, the $D_0$ gate is created using the same process as the other gates, with 10/100~nm thick Ti/Au. This gate runs over the gate $C_3$ before contacting the substrate at the centre of the dot array. The gates created in the first layer are 30 nm wide, whereas the width of the $D_0$ gate is 100 nm.

	\subsection*{Device operation and calibration of experimental parameters}
    The full set of gates $B_i$, $P_i$, $C_i$ and $D_0$ shown in Fig.~1a are designed to define and control the quantum dot plaquette. In addition, gates $X_i$, $Y_i$ and $S_i$ define two larger quantum dots which are used as charge sensors.  Different parameters of the dot array can be controlled using voltages on different gates. The $P_i$ gates are designed to control the electron filling of dot $i$ by adjusting the dot chemical potential $\mu_i$. Gates $D_0$ and $C_i$ are designed to control the tunnel coupling $t_{i,j}$, while gates $B_i$ and $C_{i+1}$ are designed to control the coupling between dot $i$ and its reservoir. In reality, the proximity between the gates causes non-negligible cross capacitances, complicating independent control of the parameters that the gates were designed to control. For some of the tuning stages, we make use of linear combinations of gate voltages--known as \emph{virtual gates}~\cite{Hensgens2017,Mukhopadhyay2018}--to provide a direct experimental knob to Hamiltonian parameters such as $\mu_i$ and $t_{i,j}$.
    
    We use charge stability diagrams~\cite{Wiel2002} to identify the charge state of the system as a function of different $P_i$ voltages. We can convert changes in gate voltages $\Delta P_i$ into changes in dot local offset energies $\Delta \mu_i$ by measuring the lever arms $\alpha_i = \Delta \mu_i / \Delta P_i$, using the method described in detail in Ref.~\cite{Hensgens2017}. For this device the measured values are $\alpha_{[1, 2, 3, 4]} = [30(2), 45(4), 55(6), 38(3)]$~$\mu$eV/mV. The uncertainty in the estimation of $\alpha_i$ is dominated by the precision with which we can identify a charge transition in the charge stability diagram, which is $\sim 1$~mV. Different features of the charge stability diagrams are also used to estimate the effective Hamiltonian parameters in the experimental system~\cite{Hensgens2017,Mukhopadhyay2018}. The effective on-site interaction $U_i$ is measured by extracting the local energy offset in dot $i$ required to change the occupation from 1 electron to 2 electrons. For this device these values where measured to be $U_{[1, 2, 3, 4]} = [2.9(2), 2.6(2), 2.9(3), 3.0(2)]$~meV. The uncertainty in the estimation of $U_i$ is calculated from the vector sum of the relative uncertainties of the $\sim 1$~mV measurement precision, and the uncertainty in $\alpha_i$ used in the conversion from voltage to energy. The effective tunnel coupling term $t_{i,j}$ is measured by analysing the width of the step in the charge sensing signal as the detuning between dots $i$ and $j$ is swept to transfer a single electron between them. For most of the results in this work, the $t_{i,j}$ terms where tuned to $16(4)$~$\mu$eV. The uncertainty in $t_{i,j}$ has roughly equal contributions from the estimation of the coupling from the fit to the width of the step, and the ability to simultaneously tune all four couplings. For the results in Fig.~4, $t_{2,3}$ was tuned to different values which are provided in the caption of the figure. The charge stability diagram simulations (Fig.~2a), require values for the interdot coulomb repulsion $V_{i,j}$ which are also extracted from measured charge stability diagrams as $V_{1,2} = 0.47(6)$~meV, $V_{2,3} = 0.35(7)$~meV, $V_{3,4} = 0.43(7)$~meV, $V_{1,4} = 0.30(4)$~meV, $V_{1,3} = 0.28(6)$~meV, $V_{2,4} = 0.18(5)$~meV.

\begin{figure}
    \centering
    \includegraphics[scale=0.8]{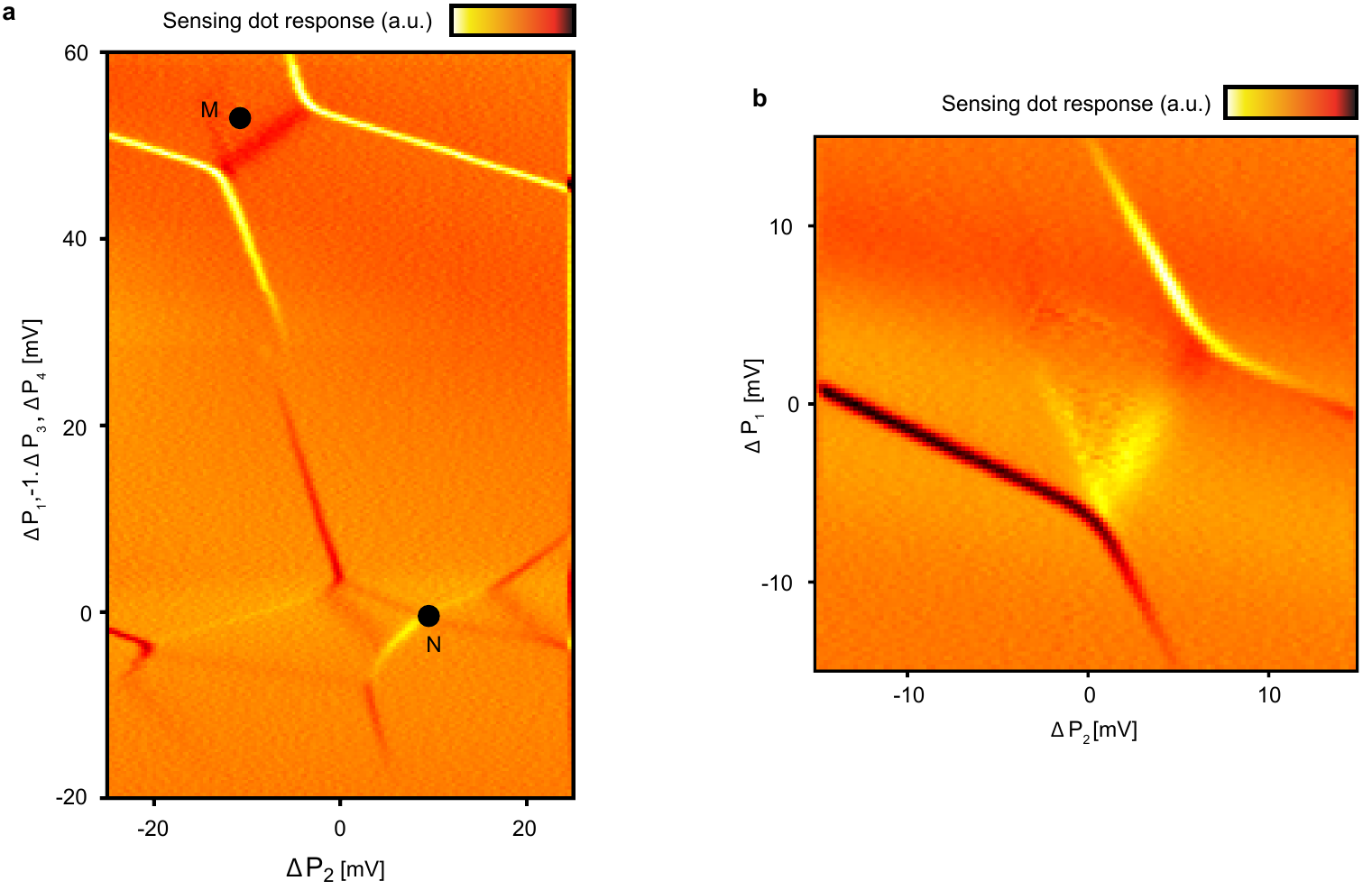} 
    \caption{\textbf{Charge stability diagram of the relevant voltage regions}. \textbf{a,} Measured charge stability diagram showing both point~$N$ and point~$M$, as highlighted in Fig.~2a. \textbf{b,} Measured charge stability diagram focusing on the 2001:1101 charge transition, where spin measurements are performed (point~$M$).\label{sfig:csd}}
    \includegraphics[scale=0.8]{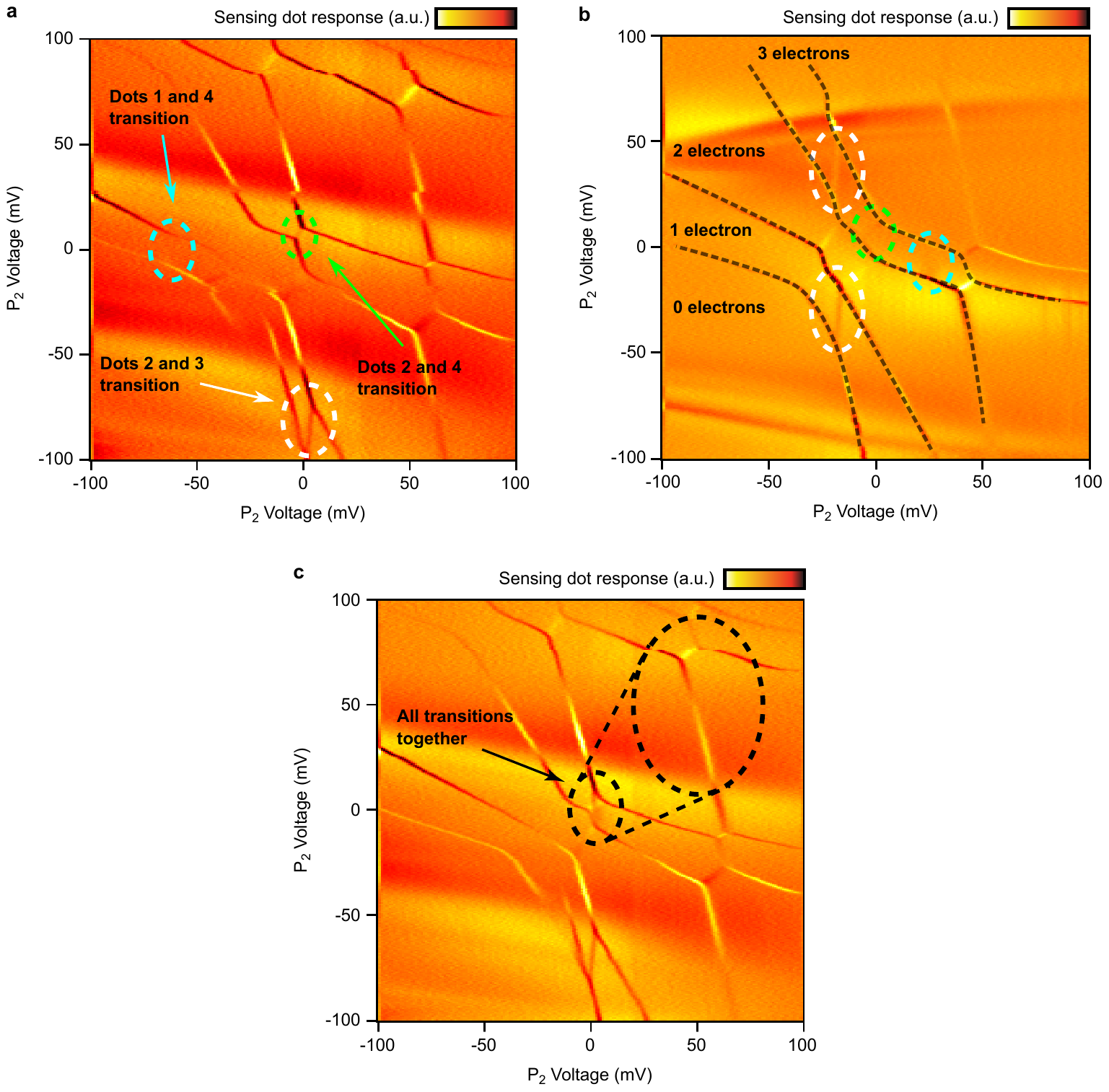} 
    \caption{\textbf{Tuning the gate voltages to the Nagaoka condition using charge stability diagrams}. \textbf{a,} Sample charge stability diagram where we have highlighted the visible interdot transitions, where the electrochemical potentials of two dots become resonant (i.e., an electron is allowed to tunnel between the two dots). \textbf{b,} Charge stability diagram similar to \textbf{a}, where we have modified the values $P_1$ and $P_3$ such that the interdot transitions appear at different locations in the diagram. Dashed black lines delimit the regions with a fixed total electron occupation in the system. \textbf{c,} In this diagram, gates $P_1$ and $P_3$ have been tuned to observe the Nagaoka condition, where the three visible interdot transitions are aligned in the three-electron configuration. The intersite interaction in the system provides an effective isolation from the reservoirs for a narrow range of gate voltages, such that the system can remain stable with three electrons in the resonant configuration.\label{sfig:align}}
\end{figure}

We make use of charge stability diagrams to observe charge tunnelling events either between an electron reservoir and a dot, or between two dots in the plaquette. These diagrams (such as the ones in Fig.~2a, Fig.~\ref{sfig:csd} and Fig.~\ref{sfig:align}) allow us to map out the charge occupation of the system as a function of voltage in the gates.
    
In order to observe signatures of Nagaoka ferromagnetism, we need to tune the system to a regime where it is loaded with 3 electrons, and the charge configuration energies of the electrons are resonant. We set the local energy reference at this regime as $\mu_i(N) = 0$~eV for all dots, and refer to this condition as point~$N$ (see Fig.~2a).
    
To tune $t_{i,i+1}$ close to point $N$, we first localise 2 of the electrons in dots $i+2$, $i+3$ (i.e. by slightly lowering $\mu_{i+2}$, $\mu_{i+3}$), and keep dots $i$, $i+1$ resonant using the remaining electron to measure their tunnel coupling. Here we use cyclic dot indices with $i = \{1,2,3,4\}$.
	
Since the sensing dots are only sensitive to charge tunnelling events, a spin-to-charge conversion protocol~\cite{Hanson2007} is needed in order to perform measurements of the spin state of the system. We do this at point~$M$, where $\mu^M_i \approx [-2.5,0.0,1.0,-0.5]$~meV (see inset of Fig.~2a). There, the ground charge state is $[2,0,0,1]$ (where $[n_1,n_2,n_3,n_4]$ corresponds to the number of electrons with dot number in the subscript), while the first excited charge state is $[1,1,0,1]$. These states have an uncoupled spin in dot 4, with the remaining 2 spins in a singlet $\lvert S \rangle$ (triplet $\lvert T \rangle$) configuration for the ground (first excited) state. We chose to use dots~$1$ and $2$ for readout, because we obtained the highest readout contrast from this pair of dots in this device.
	
The exact gate voltages required to tune the device to points $M$ and $N$ need to be calibrated visually using charge stability diagrams. In Fig.~\ref{sfig:csd}b we show a sample charge stability diagram similar to the ones used to identify the gate voltages that will tune the device to point~$M$. After the initial visual calibration, we fine-tune the gate voltages to maximise the singlet-triplet relaxation time, which was in the range of $30$ to $50$~$\mu$s in this device. We characterise the thermal excitation rate at point~$M$ by analysing the observed random telegraph signal, in which the spins spend $\sim 10\%$ to $20\%$ of the time in the triplet state, consistent with the values measured at small $p_\epsilon$ seen in the inset of Fig. 3.
	
Point~$N$ was also calibrated visually, using charge stability diagrams such as those in Fig.~\ref{sfig:align}. We note that the scale of the $t_{i,j}$ terms limit the precision with which we can identify point~$N$, since larger $t$ broadens the interdot transitions, making them harder to identify in the charge stability diagrams. 
	
Once we have fine-tuned the gate voltages at points $M$ and $N$, we define a linear combination of $P_i$ voltages that joins the two points by a straight line in gate voltage space. To do this, we define a virtual gate $VP_\epsilon$ such that a change in this gate simultaneously changes the $P_i$ gates by different amplitudes, effectively moving the system along the `detuning proportion' $p_\epsilon$ axis in Fig.~2b (see also the line along the charge stability diagram in Fig.~2a), defined such that $\mu_i(p_\epsilon) = (1-p_\epsilon) \mu^M_i$. 
	
To make sure that no unwanted charge transitions are crossed along the $p_\epsilon$ axis, we use charge stability diagrams such as those shown in Fig.~2a (simulated) and Fig.~\ref{sfig:csd} (measured), which use a gate combination that allows to see both points~$N$ and $M$ in the same diagram.

\subsection*{Measurement protocol} 
Fig.~2b presents the results of a theoretical simulation showing the lowest three multiplets of the energy spectrum of the 3-electron system, along the line that connects point~$M$ to point~$N$. Close to point~$M$ we see a typical double quantum dot spectrum corresponding to the $[2,0,0,1] \leftrightarrow [1,1,0,1]$ charge transition with the $\lvert S \rangle$ and $\lvert T \rangle$ branches, while in the region around point~$N$ the spins delocalise and we see branches corresponding to the quadruplets and doublets of the 3-electron system.
	
With this device, we can probe the spin state of the 3-electron system using the following protocol: 1 - repeatedly (10000 times) pulse rapidly from point~$N$ to point~$M$, 2 - for each repetition, perform single-shot $\lvert S \rangle$/$\lvert T \rangle$ measurements using dots $1$ and $2$ and taking 2 out of the 3 electrons, and 3 - extract the triplet probability $P_T$. Under ideal conditions, this constitutes a 2-spin projective measurement of the 3-electron system, resulting in $P_T^{(3/2)} = 1$ when the 3-electron system is in a ferromagnetic state (any of the $s = 3/2$ quadruplets). In the low-spin sector ($s = 1/2$), there are two sets of doublet states available, one of which projects 2 spins to $\lvert S \rangle$, while the other projects to $\lvert T \rangle$ (see following sections for details). In this system the doublets are effectively degenerate (see Fig.~2b), and their hybridisation will result in $P_T^{(1/2)} = 0.5$.
	
\begin{figure*}
    \includegraphics[width=\textwidth]{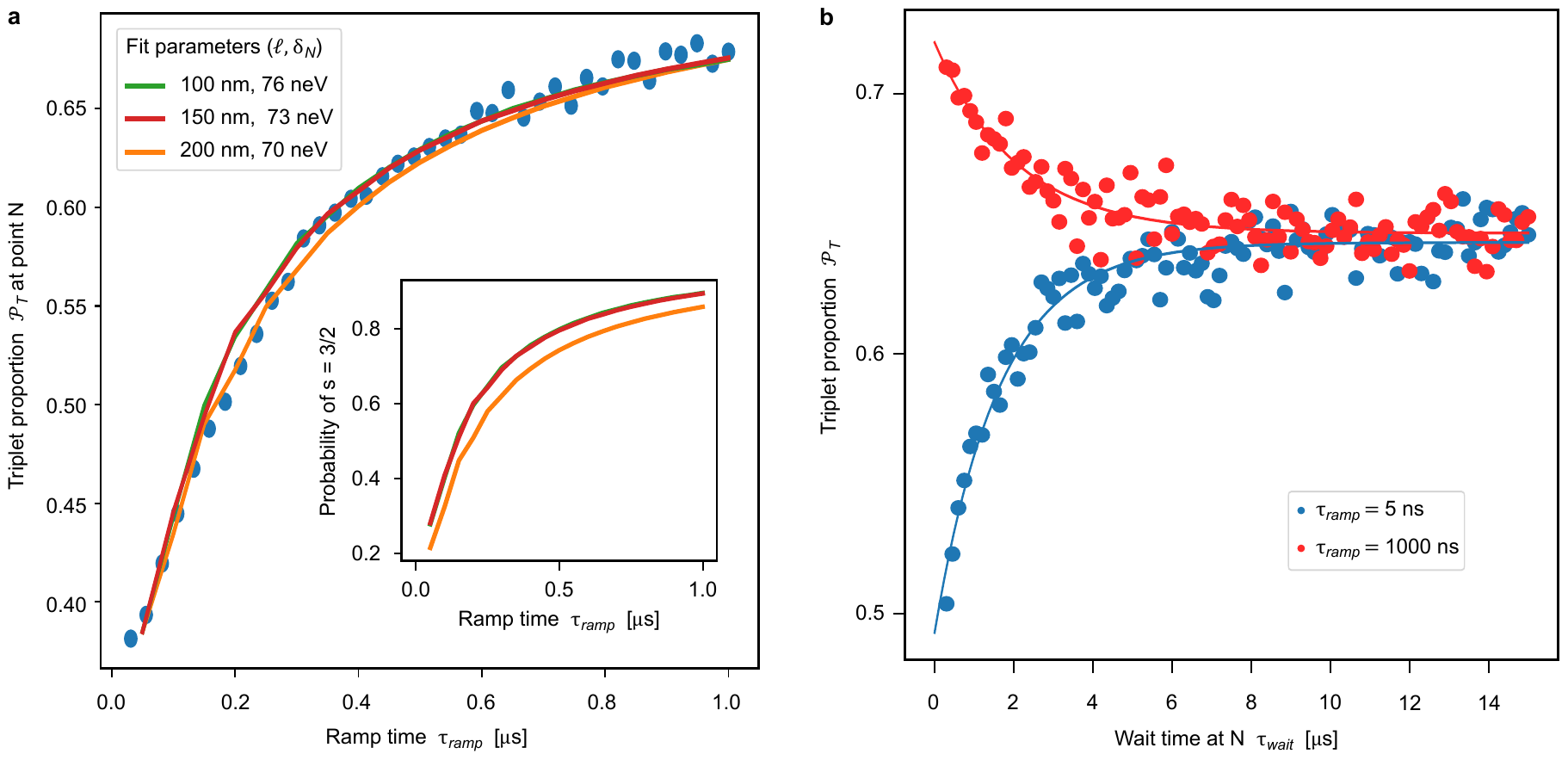}
	\caption{\textbf{Characterisation of the Nagaoka condition.} \textbf{a,} Average $P_T$ in the detuning region $1.00 < p_\epsilon < 1.01$ for $40$ values of $\tau_{ramp}$ within the same range shown in Fig.~\ref{fig:main}. Solid lines are fits using the time evolution simulations described in the supplementary material, for different values of distance $\ell$ between neighbouring dots. Inset shows the unscaled results of the time-evolution simulations, where the probability of $s = 3/2$ is the sum of the lowest 4 eigenstate probabilities from the final evolved state. \textbf{b,} Thermal relaxation measurements. $P_T$ is measured for increasing wait times at point~$N$, for diabatic (blue) and adiabatic (red) passages. Solid lines are exponential fits as guide to the eye. \label{sfig:main}}
\end{figure*}

Due to the low ratio of energy level splitting to temperature at point~$N$, we cannot probe the ground state of the system by way of relaxation. Instead, we have developed a technique similar to those previously used in quantum dot~\cite{Hanson2007} and cold atom~\cite{Trotzky2008,Nascimbene2012,Dai2017} systems, where a low-entropy state is evolved coherently to the state of interest. To do this, we apply a gate pulse sequence that follows the detuning range shown in the energy spectrum plotted in Fig.~2b. Using the pulse sequence drawn in Fig.~2c, a 2-spin singlet state with a third, free spin sitting on dot 4, is initialised by waiting at point~$M$ for $500$~$\mu$s. Next we apply a pulse on $VP_\epsilon$ towards point~$N$ of amplitude $p_\epsilon$. We then wait a time $\tau_{wait}$ at $\mu_i(p_\epsilon)$, before finally pulsing back to point~$M$ to perform the measurement. Importantly, the level crossings seen in Fig.~2b are in fact avoided level crossings with spin-orbit and nuclear hyperfine mediated coupling between the spin states (see following sections for details). This avoided level crossing allows to probe the different states in the region around $p_\epsilon = 1$, by varying the ramp rate in the pulse sequence: a slow (fast) ramp rate results in an adiabatic (diabatic) passage through the avoided level crossings, so the ground (excited) state is reached. In practice, in order to minimise leakage to excited states along the way, 80\% of the pulse is performed adiabatically, with the variable ramp time $\tau_{ramp}$ only applied to the remaining 20\%. As long as $\tau_{wait}$ is shorter than the thermal relaxation time-scale, the measurement of $P_T$ will be able to distinguish between high- and low-spin states at point~$N$. To observe relaxation of the $s = 1/2$ and $s = 3/2$ states (see Fig.~\ref{sfig:main}b), we keep $p_\epsilon = 1$ fixed and vary the wait time $\tau_{wait}$ spent at point~$N$.
	
\section*{Extended Fermi-Hubbard models used to simulate different experiments in the main text}
In this section we will describe the different parameters that are included in the model Hamiltonians that we refer to in the main text for analytical and numerical simulations. In order to capture all of the relevant effects of the experimental measurements we consider the following extended Hubbard model:
\begin{equation}
    \mathcal{H} = \mathcal{H}_H + \mathcal{H}_{so} + \mathcal{H}_{hf} + \mathcal{H}_Z
    \quad \mathrm{where} \quad
\end{equation}
\begin{equation*}
\begin{aligned}
    \mathcal{H}_H &= - \sum_{\langle j,k \rangle \sigma} t_{j,k} e^{-i\varphi_{j,k}} c^\dagger_{j\sigma} c_{k\sigma} + \sum_j U_j n_{j\uparrow} n_{j\downarrow} -\sum_j \mu_j n_j , \\
    \mathcal{H}_{so} &= \alpha(p_x\sigma_y-p_y\sigma_x)+\beta(-p_x\sigma_x+p_y\sigma_y) , \\
    \mathcal{H}_{hf} &= {\bf S}\cdot{\bf h}_N , \\
    \mathcal{H}_Z &= g\mu_B{\bf B}\cdot{\bf S}.
\end{aligned}
\end{equation*}
Each of these Hamiltonians will be described in detail in the following subsections. With the exception of the charge stability diagram simulations (which are described in the final subsection), the system is constrained to 3 electrons in the plaquette, with a maximum single-site occupation of 2 electrons, subject to Pauli exclusion (i.e. double occupation of a dot must be of opposite spin). Numerical simulations of spectra as function of {\pe} and external field were solved using the eigensolvers from the \emph{Python}-based \emph{Scipy} package. Time-evolution simulations require many iterations of matrix diagonalisation, for which we used an in-house density matrix solver package~\cite{DMsolver}.

\subsection*{Standard Fermi-Hubbard model}
The workhorse for the theoretical calculations is the standard Fermi-Hubbard model with local energy offsets
\begin{equation}
    \mathcal{H}_H = - \sum_{\langle j,k \rangle \sigma} t_{j,k} e^{-i\varphi_{j,k}} c^\dagger_{j\sigma} c_{k\sigma} + \sum_j U_j n_{j\uparrow} n_{j\downarrow} -\sum_j \mu_j n_j,
\end{equation}
where $t_{j,k}$ is the matrix element accounting for electron tunnelling between sites $j$ and $k$, $U_j$ is the on-site Coulomb repulsion energy on site $j$ and $\mu_j$ is a local energy offset at dot $j$, which can be electrostatically controlled. The operators $c_{j\sigma}$, $c_{j\sigma}^{\dagger}$ and $n_{j\sigma}$ represent the second quantisation annihilation, creation and number operators for an electron on site $j$ with spin projection $\sigma = \{ \uparrow,\downarrow \}$. The gauge $\varphi_{j,k}$ is used when applying an external magnetic field and it is described in detail in the relevant subsection below.
    
Note that for the analysis of effects related to magnetism in the system, we have omitted the intersite Coulomb interaction term, which is commonly included in extended-Hubbard models of quantum dots. In this analysis, the low-spin and high-spin states display almost identical average electron density on each site. Therefore, the long-range Coulomb interaction only lifts the total energy, but has a very small effect ($< 2$\%) on the energy gap between low- and high-spin sectors. This is further confirmed by the ab initio calculation (described in the following section), where all the interaction terms are considered and their effects are compared, showing $\sim 5$\% contribution from the long-range Coulomb terms to the low- to high-spin energy gap. The intersite term does contribute significantly to the charge state of the system in the Hubbard model, and is therefore included in the charge stability diagram simulations.

\subsection*{Spin coupling terms} 
In order to capture the {\tr} dependence of our experiments, we have added to the model the effects of spin-orbit coupling and hyperfine interactions, the two most important mechanisms that lead to spin flipping in GaAs~\cite{Stepanenko2012}.
	
For the quantum dot plaquette we have computed the matrix elements of the spin-orbit coupling Hamiltonian that accounts for the Bychkov-Rashba and the Dresselhaus effects for GaAs grown in the the crystallographic direction $[001]$:
\begin{equation}\label{seq:H1}
    \mathcal{H}_{so}=\alpha(p_x\sigma_y-p_y\sigma_x)+\beta(-p_x\sigma_x+p_y\sigma_y).
\end{equation}
here $\alpha=e\gamma_b\langle\mathcal{E}\rangle/\hbar$ and $\beta=\gamma_d\langle k_z^2\rangle/\hbar$ where $e>0$ is the elementary charge, and $\mathcal{E}$ is the electric field at the interface of the structure. For GaAs $\gamma_b\approx5.2\times10^{-2}\,\textrm{nm}^2$ and $\gamma_d\approx27.6\,\textrm{meV.nm}^3$~\cite{Winkler2003}. The axes of the coordinate system $x$ and $y$ correspond to the directions $[100]$ and $[010]$. When spin-orbit coupling is weak we may take as a basis the Wannier states $|j\rangle$ that are localised on the dots indexed by $j$. In this basis the matrix elements of \eref{seq:H1} are
\begin{equation}\label{seq:melem}
    \langle j|\mathcal{H}_{so}|k\rangle=\alpha (p_x^{jk}\sigma_y-p_y^{jk}\sigma_x)+\beta(-p_x^{jk}\sigma_x+p_y^{jk}\sigma_y),
\end{equation}
where $p_a^{jk}=\langle j|p_a|k\rangle$, $a=x,y$. Those matrix elements vanish if $j=k$. Then in the second quantised form \eref{seq:H1} reads
\begin{equation}\label{seq:H2}
    \mathcal{H}_{so}=\sum_{jk\sigma\sigma'}c_{j\sigma}^\dagger\omega^{jk}\cdot\sigma^{\sigma\sigma'}c_{k\sigma'},
\end{equation}
with $\omega^{jk}\cdot\sigma^{\sigma\sigma'}=(-\alpha p_y^{jk}-\beta p_x^{jk})\sigma_{x}^{\sigma\sigma'}+(\alpha p_x^{jk}+\beta p_y^{jk})\sigma_{y}^{\sigma\sigma'}$. 
The unit vector in the direction of the dots $j$ and $k$ is $\hat{\ell}_{jk}=\cos(\theta_{jk})\hat{x}+\sin(\theta_{jk})\hat{y}$. Eliminating the matrix elements of the momentum in the direction perpendicular to $\hat{\ell}_{jk}$, \eref{seq:H2} becomes
\begin{equation}
    \mathcal{H}_{so} = \sum_{jk\sigma\sigma'}c_{j\sigma}^\dagger p_{\ell}^{jk}\big((-\alpha\sin(\theta_{jk})-\beta\cos(\theta_{jk}))\sigma_x^{\sigma\sigma'} +(\alpha\cos(\theta_{jk})+\beta\sin(\theta_{jk})) \sigma_y^{\sigma\sigma'}\big)c_{k\sigma'}.
\end{equation}
Here $p_{\ell}^{jk}=m\langle j|\dot{\ell}|k\rangle = i m t_{jk}\ell_{jk}/\hbar$, where $m$ is the effective mass of the electron, $\ell_{jk}=\ell_j-\ell_k$ with $\ell_j$ the coordinate of dot $j$ on the $(jk)$ axis, and $t_{jk}$ equals minus the matrix element of the one-electron Hamiltonian. Therefore
\begin{equation}\label{seq:Hfin}
    \mathcal{H}_{so}=\sum_{\langle j,k \rangle }t_{jk}c_{j\uparrow}^\dagger\big(\frac{\ell_{jk}}{\lambda_b}e^{-i\theta_{jk}}-i\frac{\ell_{jk}}{\lambda_d}e^{i\theta_{jk}}\big)c_{k\downarrow}+h.c.,
\end{equation}
where $jk$ are restricted to neighbouring dots and we define the length scales $\lambda_b=\hbar/m\alpha$ and $\lambda_d=\hbar/m\beta$. Typically $\langle k_z^2\rangle\sim 0.02\,\textrm{nm}^{-2}$ and $\langle e\mathcal{E}\rangle\sim 3\,\textrm{meV/nm}$. So $\lambda_b\approx7\,\mu\textrm{m}$ and $\lambda_d\approx2\,\mu\textrm{m}$, for neighbour quantum dots ($\ell_{jk}\approx0.15\mu\textrm{m}$), giving $\ell_{jk}/\lambda_b\sim 0.02$ and $\ell_{jk}/\lambda_d\sim 0.08$.
	
The large abundance of nuclear spins in the GaAs crystal means that each site in the plaquette will be hyperfine coupled to a number of randomly oriented nuclear spins, causing each site to experience a slightly different Overhauser field. This interaction is described by the hyperfine coupling Hamiltonian$^{24,44,45}$
\begin{equation}\label{seq:HN}
    \mathcal{H}_{hf}={\bf S}\cdot{\bf h}_N.
\end{equation}
Here ${\bf S}=(\sigma_x,\sigma_y,\sigma_z)/2$ is the electron spin operator, ${\bf h}_N=\sum_{i}A_i{\bf I}_i$, $A_i=Av_0|\psi({\bf r}_i)|^2$ is the coupling parameter with nucleus $i$ having spin operator ${\bf I}_i$, $\psi({\bf r}_i)$ is the electron envelope wave function at the nuclear site ${\bf r}_i$, and $v_0$ is the volume of the crystal cell. Hence ${\bf B}_N={\bf h}_N/g\mu_B$ is the nuclear magnetic field acting on the electron with g-factor $g$, and $\mu_B$ is the Bohr magneton. 
    
The classical probability distribution of $h_{Na}$ ($a=x,y\text{ or }z$) is normal:$^{44,45}$ $P(h_{Na})=\frac{1}{\sqrt{2\pi\delta_N^2}}\exp(-h_{Na}^2/2\delta_N^2)$. The typical magnitude of the field component is $\delta_N\sim A/\sqrt{N}\ll h_{N\text{max}}\sim A$, with $N$ the number of nuclei covered by the envelope function of the electron and $h_{N\text{max}}$ the magnitude of the field when the nuclear spins are fully polarised.
For GaAs: $N\sim 10^{6}$ and $B_{N\text{max}}/\sqrt{N}$ is of the order of a few mT,$^{24}$ hence $h_{N\text{max}}/\sqrt{N}\sim0.1\,\mu{\rm eV}$. 
    
\begin{figure}
    \includegraphics[width=\textwidth]{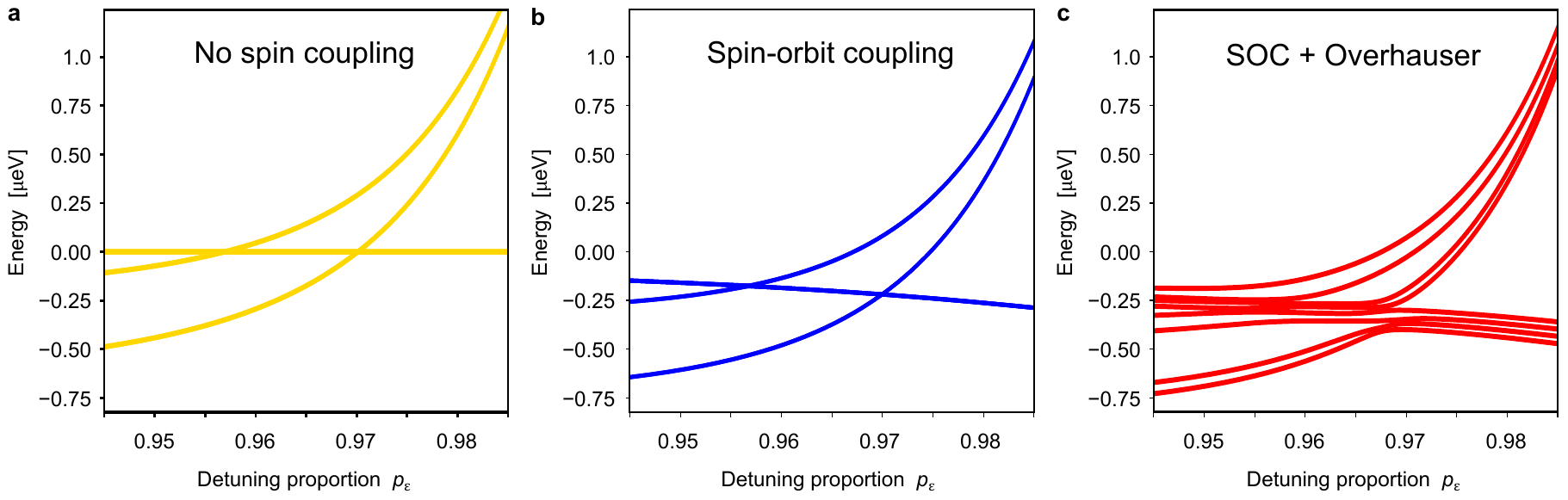}
	\caption{\textbf{Effects of spin-coupling mechanisms.} Calculated spectra of the system in the region of $p_\epsilon$ close to the level crossing of the $s = 1/2$ and $s = 3/2$ energies, comparing the effects of different mechanisms for spin coupling: \textbf{a,} Spectrum without any spin coupling effects; \textbf{b,} Spectrum including only spin-orbit coupling effects; \textbf{c,} Sample spectrum with both spin-orbit and hyperfine induced Overhauser field gradients, using a single combination of $h_{Na}$ fields selected from a normal distribution with standard deviation $\delta_N = 73$~neV. The supplementary material describe the implementations of these spin-coupling terms in the theoretical model. \label{sfig:crossing}}
\end{figure}

Since our basis states are eigenstates of the Pauli matrix $\sigma_z$, we express \eref{seq:HN} as:
\begin{equation}\label{seq:HN2}
    \mathcal{H}_{hf}=\frac{1}{2}\left(\sigma_zh_{Nz}+\sigma_{+}\left(h_{Nx}-ih_{Ny}\right)+\sigma_{-}\left(h_{Nx}+ih_{Ny}\right)\right),
\end{equation}
where $\sigma_\pm=(\sigma_x\pm i\sigma_y)/2$. We numerically implement \eref{seq:HN2} and the nuclear fields of the four quantum dots are taken to be independent. In Fig.~\ref{sfig:crossing} we show that the effect of the hyperfine coupling dominates over spin-orbit coupling, in the detuning region of the energy level crossings.
    
\subsection*{External magnetic field} 
To capture the orbitals effects resulting from a magnetic flux through the square plaquette, we included a Peierls phase $\varphi_{jk}$ to the tunneling matrix elements of $\mathcal{H}_H$:
\begin{equation}
    \varphi_{jk}=\frac{e}{\hbar}\int_{{\bf r}_k}^{{\bf r}_j}d{\bf r}\cdot{\bf A}({\bf r})=\frac{2\pi}{\Phi_0}\int_{{\bf r}_k}^{{\bf r}_j}d{\bf r}\cdot{\bf A}({\bf r}),
\end{equation}
where $e>0$ is the elementary charge, $\hbar$ is the reduced Planck constant, $\Phi_0=h/e$ is the flux quantum, and ${\bf A}({\bf r})$ is the magnetic vector potential. We use the gauge for which $\varphi_{41}=2\pi\Phi/\Phi_0$, with $\Phi=B\ell^2$ the magnetic flux through the plaquette and $\ell$ the length of the side of the plaquette, and the phases for the other links vanish.
    
The Zeeman contribution is:
\begin{equation}
    \mathcal{H}_Z=g\mu_B{\bf B}\cdot{\bf S},
\end{equation}
where ${\bf B}$ is the external magnetic field.
	
\subsection*{Representation of the quantum states} 
We now describe our methodology for constructing the Hamiltonian matrices. For the 2$\times$2 plaquette:
\begin{equation}
     |\psi\rangle=\sum_{\{n_{i\sigma}\}}a(\{n_{i\sigma}\})|\{n_{i\sigma}\}\rangle.
\end{equation}
The basis consists of the states specified by the occupations of the electrons on the lattice and their spin projections: 
\begin{equation}\label{number}
    |\{n_{i\sigma}\}\rangle=|n_{1\uparrow}n_{2\uparrow}n_{3\uparrow}n_{4\uparrow}n_{1\downarrow}n_{2\downarrow}n_{3\downarrow}n_{4\downarrow}\rangle,
\end{equation}
with $n_{i\sigma}=0\text{ or }1$. For $N$ electrons on the plaquette we have $\sum_{i\sigma}n_{i\sigma}=N$ and the basis states consist of all combinations of the occupations at fixed $N$. Hence $N=3$ corresponds to a space of the quantum states of dimension $8!/5!3!=56$.
    
The on-site energy and the Coulomb repulsion terms of the Hamiltonian $\sum_i U_i n_{i\uparrow} n_{i\downarrow} - \sum_i \mu_i n_i$ are diagonal in this basis. Tunnelling involves the off-diagonal matrix elements~\cite{Landau2013}: 
\begin{equation}
    \langle\dots1_{i\sigma}\dots0_{j\sigma'}\dots|c_{i\sigma}^\dagger c_{j\sigma'}|\dots0_{i\sigma}\dots1_{j\sigma'}\dots\rangle=(-1)^{\Sigma_{\ell=i\sigma+1}^{j\sigma'-1}n_\ell},
\end{equation}
where $\ell$ goes over the elements between $i\sigma$ and $j\sigma'$ (exclusive) in the list (\ref{number}). 
The Hamiltonian commutes with the spin operators ${\bf S}^2$ and $S_z$ and its eigenstates are also spin eigenstates $|s,m\rangle_\alpha$:
\begin{equation}
\begin{aligned}
    {\bf S}^2|s,m\rangle_\alpha&=s(s+1)|s,m\rangle_\alpha,\\
    S_z|s,m\rangle_\alpha&=m|s,m\rangle_\alpha,\,m=-s,-s+1,\dots,s.
\end{aligned}
\end{equation}
The spin operators are ${\bf S}^2=S_z^2+\frac{1}{2}(S_+S_-+S_-S_+)$, $S_z=\frac{1}{2}\sum_i(n_{i\uparrow}-n_{i\downarrow})$, $S_+=\sum_ic_{i\uparrow}^\dagger c_{i\downarrow}$, and $S_-=\sum_ic_{i\downarrow}^\dagger c_{i\uparrow}$. The label $\alpha$ distinguishes between the states with the same quantum numbers $s$ and $m$. For three electrons in the absence of a magnetic field those states consist of energy degenerate $s=3/2$ quadruplets and two sets of energy degenerate $s=1/2$ doublets. In the low-energy sector relevant to the study, $\alpha$ distinguishes between the two sets of $s=1/2$ doublets.
	
\subsection*{Basic construction of the 3-electron filled plaquette Fermi-Hubbard Hamiltonian} 
For the initial characterisation of Nagaoka ferromagnetism in a 2$\times$2 plaquette, we consider only $\mathcal{H}_H$ with homogeneous interactions ($U_i = U$, $t_{i,j} = t$, $\mu_i = 0$) and no external field ($\varphi_{i,j} = 0$). In this simplest configuration, the model can be solved analytically. The Hamiltonian can be divided into two independent blocks, one for the $m = \pm3/2$ states (parallel spins) and another for the $m = \pm1/2$ states (one flipped spin):
\begin{equation}
	\bm{\mathcal{H}_H} = \bm{\mathcal{H}_{3/2}} + \bm{\mathcal{H}_{1/2}}
\end{equation}
and for each block it is sufficient to solve for one of the $m$ projections and assume another degenerate set of states for the opposite $m$ projection. As will be shown, these assumptions reduce the dimensions of the Hamiltonians to $4$ for $\bm{\mathcal{H}_{3/2}}$ and $24$ for $\bm{\mathcal{H}_{1/2}}$, making them simpler to solve analytically.
	
The quantum states for $\bm{\mathcal{H}_{3/2}}$ will be 
\begin{equation}
	\ket{\psi_{3/2}} = a_1 \ket{0 \uparrow \uparrow \uparrow} + a_2 \ket{\uparrow 0 \uparrow \uparrow} + a_3 \ket{\uparrow \uparrow 0 \uparrow} + a_4 \ket{\uparrow \uparrow \uparrow 0}
\end{equation}
with the Hamiltonian
\begin{equation}
	\bm{\mathcal{H}_{3/2}} = 
	\begin{bmatrix}
		0 & -t & 0 & -t\\
		-t & 0 & -t & 0\\
		0 & -t & 0 & -t\\
		-t & 0 & -t & 0
	\end{bmatrix}
\end{equation}
with eigenvalues $\{-2t, 0, 2t\}$.

For the block with $m = \pm1/2$, double occupation is allowed, therefore we need to consider more available states. We construct the Hamiltonian by first fixing the flipped spin in one dot and working out all the possible states in the basis. For example, a down spin in dot 1 results in the basis sub-set 
\begin{equation}
    \ket{\psi'_{1/2}} = a_1 \ket{2 \uparrow 0 0} + a_2 \ket{2 0 \uparrow 0} + a_3 \ket{2 0 0 \uparrow} + a_4 \ket{\downarrow 0 \uparrow \uparrow} + a_5 \ket{\downarrow \uparrow 0 \uparrow} + a_6 \ket{\downarrow \uparrow \uparrow 0}
\end{equation}
from which we then construct
\begin{equation}
	\bm{\mathcal{H}'_{1/2}} = 
	\begin{bmatrix}
		U  & -t & 0  & 0  & t  & 0 \\
		-t & U  & -t & -t & 0  & t \\
		0  & -t & U  & 0  & -t & 0 \\
		0  & -t & 0  & 0  & -t & 0 \\
		t  & 0  & -t & -t & 0  & -t\\
		0  & t  & 0  & 0  & -t & 0
	\end{bmatrix}
\end{equation}
The same matrix can be used for the subspace with the flipped spin on each of the other dots. To finish constructing the 24-dimensional Hamiltonian, we need to then work out the hopping matrices for the spin down, which results in the full Hamiltonian:
\begin{equation}
	\bm{\mathcal{H}_{1/2}} = 
	\begin{bmatrix}
		\bm{\mathcal{H}'_{1/2}} & \bm{T} & 0 & \bm{T^{\intercal}}\\
		\bm{T^{\intercal}} & \bm{\mathcal{H}'_{1/2}} & \bm{T} & 0\\
		0 & \bm{T^{\intercal}} & \bm{\mathcal{H}'_{1/2}} & \bm{T}\\
		\bm{T} & 0 & \bm{T^{\intercal}} & \bm{\mathcal{H}'_{1/2}}
	\end{bmatrix}
	\quad \mathrm{where} \quad
	\bm{T} =
	\begin{bmatrix}
		0  & 0  & t  & 0  & 0  & 0 \\
		0  & 0  & 0  & 0  & t  & 0 \\
		0  & 0  & 0  & 0  & 0  & t \\
		-t & 0  & 0  & 0  & 0  & 0 \\
		0  & -t & 0  & 0  & 0  & 0 \\
		0  & 0  & 0  & -t & 0  & 0
	\end{bmatrix}
\end{equation}
The lowest two eigenvalues of this Hamiltonian are $-2t$ (two states) and  $-\sqrt{3} t - \frac{5t^2}{U}\}$ (four states). The former correspond to the $m = \pm1/2$ states of the quadruplets with total spin $s = 3/2$. The remaining four states correspond to the two $s = 1/2$ doublets (with $m = \pm1/2$).
	
As expected for a 3-spin system, the 8 lowest eigenenergies of this Hamiltonian contain 4 degenerate ferromagnetic quadruplets and the 2 sets of degenerate low-spin doublets.
	
For detuning spectra simulations, we construct the Hamiltonian similarly as above but with inhomogeneous $U_i$, $t_{i,j}$ and $\mu_i$ parameters to reproduce the experimental conditions, and extract the eigenenergies numerically. The values of $U_{[1, 2, 3, 4]} = [2.9, 2.6, 2.9, 3.0]$~meV where the same previously measured in this device.$^{4}$ When considering spin-coupling terms, it is necessary to use the full quantum state representation with the 56-dimensional Hilbert space.
	
\subsection*{Time evolution simulations}
We use time-evolution simulations to extract information about the spin-coupling mechanisms at the avoided crossings from the results in Fig. 4a. We use the full Hamiltonian with spin-coupling to simulate the conditions in the $p_\epsilon$ pulsing experiments. In the experiment, we initialise the ground state at point~$M$ and ramp adiabatically to $p_\epsilon = 0.8$, before pulsing to point~$N$ with a variable ramp time~$\tau_{ramp}$.
	
We use an in-house solver package~\cite{DMsolver} to simulate the evolution of the initialised state for the last 20$\%$ of $p_\epsilon$ with varying ramp times. At $p_\epsilon = 0.8$, we consider the initialised state as a statistical mixture of the two lowest energy eigenstates, both of which are $s = 1/2$ states at $p_\epsilon = 0.8$. We consider 20 values of $\tau_{ramp}$ in the range from 50~ns to 1~$\mu$s, taking 10000 time-steps for each ramp. We then add the overlaps of the averaged density matrix with each of the four lowest energy eigenstates at point~$N$ (i.e., the eigenstates with $s = 3/2$). This overlap can be mapped to an ideal $P_T$ measurement with the method described two sections below. For each value of $\tau_{ramp}$, we repeat the evolution 350 times, drawing different values of $h_{Na}$, and compute the average $P_T$ for the final state. To account for imperfections of the experimental measurement of $P_T$--caused by the finite measurement bandwidth, the signal to noise ratio and {\trip} to {\sing} relaxation, as well as unwanted leakage to other states during the pulsed passages--we scale the ideal calculated values of $P_T(\tau_{ramp})$ to match the experimental $P_T$ at the minimum and maximum value of {\tr}.
	
We vary the parameter $\delta_N$ and use the method above to get the best fit to our experimental data. Additionally, the spin-orbit term requires an estimate of the distance between neighbouring dots, which was lithographically designed to be $\ell = 150$~nm. We consider a conservative range of $\ell$ from $100$ to $200$~nm (see Fig. 4a), from which we extract the estimate for $\delta_N = 73 \pm 3$~neV quoted in the main text. Previous observations and calculations of this parameter in similar GaAs quantum dot systems$^{46-48}$ have estimated it to be in the range of $70$~neV to $120$~neV.

\subsection*{Extracting $\delta_N$ using the Landau-Zener model} 
The nuclear fields lead to the lifting of the spin degeneracies of the $s=3/2$ quadruplet and the $s=1/2$ doublets and multiple avoided crossings of the order of $\delta_N$. A simple estimate of the characteristic time-scale of crossover between the diabatic and the adiabatic regimes of voltage tuning can be obtained by using the Landau-Zener formula for a two-level system~\cite{Landau2013}. Then the characteristic ramp time is 
\begin{equation}\label{seq:LZ}
    \tau_{ramp}^*=\frac{\hbar \Delta p_\epsilon}{2\pi \delta_N^2}\frac{d\Delta E}{dp_\epsilon}.
\end{equation}
For $\Delta p_\epsilon=0.2$ this gives $\tau_{ramp}^*\sim100\,{\rm ns}$, which is consistent with the time scale obtained by the time-dependent numerical simulation of the model.
	
\subsection*{Charge stability simulations} 
In this work we use charge stability diagrams to identify different charge occupation regimes and charge transitions as function of gate voltages. Simulation of charge stability diagrams requires a slightly modified version of the theoretical model. We use $\mathcal{H}_H$ with the addition of an intersite Coulomb repulsion term $\sum_{i<j} V_{i,j} n_i n_j$, with $V_{1,2} = 0.47$, $V_{2,3} = 0.35$, $V_{3,4} = 0.43$, $V_{1,4} = 0.30$, $V_{1,3} = 0.28$, $V_{2,4} = 0.18$, previously measured in this device$^{4}$. The number of basis states is expanded such that the total occupation of the system can vary from $0$ to $2$ electrons per site. Additionally, we use gate to local energy lever arms and a cross-capacitance matrix measured from experiment to implement gate voltages $P_i$ into the model and calculate their effect on local energies $\mu_i$. We use this model to calculate charge occupation as a function of gate voltages. The Hamiltonians are constructed and solved the simulation toolbox in the Python based package \emph{qtt}~\cite{qtt}.
		
\section*{Mapping 3-spin states onto 2-spin measurements}
In the main text, we state that we can distinguish between the 3-spin $s = 1/2$ and $s = 3/2$ states through a projective singlet/triplet ({\sing/\trip}) measurement on 2 random spins. Here we show this in the first-quantisation formulation of the spin states. We use the following 8 basis states of the system with 3 spin-$\frac{1}{2}$ particles: 
\begin{equation}
\begin{aligned}
    \ket{\frac{3}{2}, +\frac{3}{2}} &= \ket{\uparrow \uparrow \uparrow}\\
    \ket{\frac{3}{2}, +\frac{1}{2}} &= \frac{1}{\sqrt{3}} \left( \ket{\uparrow \uparrow \downarrow} + \ket{\uparrow \downarrow \uparrow} + \ket{\downarrow \uparrow \uparrow} \right)\\
    \ket{\frac{3}{2}, -\frac{1}{2}} &= \frac{1}{\sqrt{3}} \left( \ket{\downarrow \downarrow  \uparrow } + \ket{\downarrow \uparrow  \downarrow} + \ket{ \uparrow \downarrow \downarrow} \right)\\
    \ket{\frac{3}{2}, -\frac{3}{2}} &= \ket{\downarrow \downarrow \downarrow}\\
    \ket{\frac{1}{2}, +\frac{1}{2}}_1 &= \frac{1}{\sqrt{3}} \left(- \ket{\uparrow \uparrow\downarrow } + e^{i\pi/3}\ket{ \uparrow \downarrow \uparrow} +  e^{-i\pi/3}\ket{\downarrow \uparrow  \uparrow} \right)\\
     \ket{\frac{1}{2}, -\frac{1}{2}}_1 &= \frac{1}{\sqrt{3}} \left(- \ket{\downarrow\downarrow \uparrow } + e^{i\pi/3}\ket{\downarrow \uparrow  \downarrow} +  e^{-i\pi/3}\ket{\uparrow\downarrow   \downarrow} \right)\\
    \ket{\frac{1}{2}, +\frac{1}{2}}_2 &= \frac{1}{\sqrt{3}} \left(- \ket{\uparrow \uparrow\downarrow } + e^{-i\pi/3}\ket{ \uparrow \downarrow \uparrow} +  e^{i\pi/3}\ket{\downarrow \uparrow  \uparrow} \right)\\
    \ket{\frac{1}{2}, -\frac{1}{2}}_2 &= \frac{1}{\sqrt{3}} \left(- \ket{\downarrow\downarrow \uparrow } + e^{-i\pi/3}\ket{\downarrow \uparrow  \downarrow} + e^{i\pi/3} \ket{\uparrow \downarrow \downarrow} \right)
\end{aligned}
\end{equation}
    
The 2-spin system has one singlet ($\ket{S}$) and three triplet states ($\ket{T_+}, \ket{T_0}, \ket{T_-}$), given by:
\begin{equation}
\begin{aligned}
    &\ket{S} = \ket{0, 0} = \frac{1}{\sqrt{2}} \left( \ket{\uparrow \downarrow} - \ket{ \downarrow \uparrow} \right) \\
    &\ket{T_+} = \ket{1, +1} =  \ket{\uparrow \uparrow} \\
    &\ket{T_0} = \ket{1, 0} = \frac{1}{\sqrt{2}} \left( \ket{\uparrow \downarrow} + \ket{ \downarrow \uparrow} \right) \\
    &\ket{T_-} = \ket{1, -1} =  \ket{\downarrow \downarrow}
\end{aligned}
\end{equation}    
To obtain the 2-spin projection on the 3-spin system, we take partial inner products of each of the eight basis  states with singlet and triplet states in the first two spins. First, we take the basis state $ \ket{\frac{3}{2}, +\frac{3}{2}}$ :    
\begin{equation}
\begin{aligned}
    \braket{S}{\frac{3}{2}, +\frac{3}{2}} &= \frac{1}{\sqrt{2}} \left[ \bra{\uparrow\downarrow} - \bra{\downarrow\uparrow} \right] \left[\ket{\uparrow \uparrow \uparrow} \right] = 0\\
    \braket{T_0}{\frac{3}{2}, +\frac{3}{2}} &= \frac{1}{\sqrt{2}} \left[ \bra{\uparrow\downarrow} + \bra{\downarrow\uparrow} \right] \left[\ket{\uparrow \uparrow \uparrow} \right] = 0\\
    \braket{T_+}{\frac{3}{2}, +\frac{3}{2}} &= \bra{\uparrow\uparrow}  \ket{\uparrow \uparrow \uparrow}  = \ket{\uparrow}\\
    \braket{T_+}{\frac{3}{2}, +\frac{3}{2}} &= \bra{\downarrow\downarrow}  \ket{\uparrow \uparrow \uparrow}  = 0
\end{aligned}
\end{equation}    
The probability of a {\sing} measurement outcome in a 2-spin projective measurement of the $ \ket{\frac{3}{2}, +\frac{3}{2}}$ basis state is $\lVert\braket{S}{\frac{3}{2}, +\frac{3}{2}}\rVert^2 = 0 $. Similarly, the probability of a {\trip} measurement outcome is
\begin{equation*}
    \lVert\braket{T_+}{\frac{3}{2}, +\frac{3}{2}}\rVert^2 + \lVert\braket{T_0}{\frac{3}{2}, +\frac{3}{2}}\rVert^2 + \lVert\braket{T_-}{\frac{3}{2}, +\frac{3}{2}}\rVert^2 = 1 + 0 + 0 = 1.
\end{equation*}
Following similar derivations, we find that also for the other three basis states with $s = 3/2$, the probabilities of obtaining {\sing} and {\trip} upon measurement are 0 and 1 respectively.
	
Next, we take the basis state $\ket{\frac{1}{2}, +\frac{1}{2}}_1$ :     
\begin{equation}
\begin{aligned}
    \braket{S}{\frac{1}{2}, +\frac{1}{2}}_1 &= \frac{1}{\sqrt{2}} \left[ \bra{\uparrow\downarrow} - \bra{\downarrow\uparrow} \right] \frac{1}{\sqrt{3}} \left[- \ket{\uparrow \uparrow\downarrow } + e^{i\pi/3}\ket{ \uparrow \downarrow \uparrow} +  e^{-i\pi/3}\ket{\downarrow \uparrow  \uparrow} \right]\\
    &= \frac{1}{\sqrt{6}} \left[ e^{i\pi/3} -  e^{-i\pi/3}\right] \ket{\uparrow} = \frac{i}{\sqrt{2}} \ket{\uparrow}\\
    \braket{T_0}{\frac{1}{2}, +\frac{1}{2}}_1 &= \frac{1}{\sqrt{2}} \left[ \bra{\uparrow\downarrow} + \bra{\downarrow\uparrow} \right] \frac{1}{\sqrt{3}} \left[- \ket{\uparrow \uparrow\downarrow } + e^{i\pi/3}\ket{ \uparrow \downarrow \uparrow} +  e^{-i\pi/3}\ket{\downarrow \uparrow  \uparrow} \right] \\
    &= \frac{1}{\sqrt{6}} \left[ e^{i\pi/3} +  e^{-i\pi/3}\right] \ket{\uparrow} = \frac{1}{\sqrt{6}} \ket{\uparrow}\\
    \braket{T_+|\frac{1}{2}, +\frac{1}{2}}_1 &= \bra{\uparrow\uparrow}  \frac{1}{\sqrt{3}} \left[- \ket{\uparrow \uparrow\downarrow } + e^{i\pi/3}\ket{ \uparrow \downarrow \uparrow} +  e^{-i\pi/3}\ket{\downarrow \uparrow  \uparrow} \right]  = - \frac{1}{\sqrt{3}}\ket{\downarrow}\\
    \braket{T_-}{\frac{1}{2}, +\frac{1}{2}}_1 &= \bra{\downarrow\downarrow}  \frac{1}{\sqrt{3}} \left[- \ket{\uparrow \uparrow\downarrow } + e^{i\pi/3}\ket{ \uparrow \downarrow \uparrow} +  e^{-i\pi/3}\ket{\downarrow \uparrow  \uparrow} \right]  = 0
\end{aligned}
\end{equation}
This results in 2-spin measurement probabilities of:
\begin{equation*}
\begin{aligned}
    &\lVert\braket{S}{\frac{1}{2}, +\frac{1}{2}}_1\rVert^2 = \frac{1}{2} 
    \quad \mathrm{and}\\
    &\lVert\braket{T_+}{\frac{1}{2}, +\frac{1}{2}}_1\rVert^2 + \lVert\braket{T_0}{\frac{1}{2}, +\frac{1}{2}}_1\rVert^2 + \lVert\braket{T_-}{\frac{1}{2}, +\frac{1}{2}}_1\rVert^2 = \frac{1}{3} + \frac{1}{6} + 0 = \frac{1}{2}.
\end{aligned}
\end{equation*}
Similar calculations for the other three basis states with $s = 1/2$ show {\sing} and {\trip} measurement probabilities of $0.5$ each. Although we have used the 2 spin singlet and triplet states for the first two spins for the calculations, same results hold for any other two spin combinations.\\
Assuming statistical mixing of the 8 basis states with 3 spin-$1/2$ particles, the probability of a two-spin singlet measurement outcome is given by:
\begin{equation*}
    P_S=\sum_{s,m}P(s,m)||\langle S|s,m\rangle||^2
\end{equation*}
where $P(s,m)$ is the probability of occupation of the three-electron spin state $|s,m\rangle$. Similarly the probability of a two-spin triplet measurement outcome is given by:
\begin{equation*}
    P_T=\sum_{s,m}P(s,m)\left[||\langle T_+|s,m\rangle||^2 + ||\langle T_0|s,m\rangle||^2 + ||\langle T_-|s,m\rangle||^2 \right]
\end{equation*}
As we have seen before, for any basis state with $s = 3/2$, the probability two-spin triplet measurement outcome is 1. So, for any statistical mixture of $s = 3/2$ basis states, the probability a two-spin triplet measurement outcome is also 1. Similarly, for any statistical mixture of $s = 1/2$ basis states, the probability a two-spin triplet measurement outcome is $0.5$. So in our experiment the expected values of $P_T^{3/2}$ and $P_T^{1/2}$ are 1 and 0.5, where $P_T^{3/2}$ ($P_T^{1/2}$) is the probability a two-electron triplet state measurement outcome from the quadruplet (doublet) configuration.

\section*{Ab initio exact diagonalisation simulations of the 2$\times$2 plaquette}
We have developed an ab initio model of the quantum dot plaquette used in the experiments,$^{41}$ in order to provide, in some aspects, more realistic benchmarks than the single-band Hubbard model used throughout. This calculation employs the many-body wavefunction bases constructed by the Slater determinant of eigentstate of Gaussian quantum wells $V(\mathbf{r}) = -V_0 e^{-|\mathbf{r}|^2/2\delta}$. In the second-quantized form of the bases, the quadratic part of the Hamiltonian is 
\begin{equation}
	h_{i\alpha,j\beta} = \expval{i \alpha \sigma}{\mathcal{H}}{j \beta \sigma},
\end{equation}
where $h_{i\alpha,j\beta}$ define the site energy (diagonal terms) and hybridisation (off-diagonal terms) of the orthonormal orbitals, for dot centres $\{i,j\}$, and orbital and spin indexes $\{\alpha,\beta\}$ and $\sigma$ respectively. The on-site interactions are computed after formulating an on-site multiplet model~\cite{Dagotto2003}:
\begin{equation}
\begin{aligned}
	\Ham^{\rm (int)}_i &= \frac12\sum_{\alpha\sigma} U_{\alpha} n_{\alpha\bar{\sigma}}n_{\alpha\sigma} + \frac12\!\sum_{\alpha_1\neq\alpha_2}\!\sum_{\sigma_1,\sigma_2} U^\prime_{\alpha_1 \alpha_2} n_{\alpha_2\sigma_2}n_{\alpha_1\sigma_1}\\
    &\quad + \frac12\sum_{\alpha_1\neq\alpha_2}\sum_{\sigma_1,\sigma_2} J_{\alpha_1 \alpha_2}c^\dagger_{\alpha_2\sigma_1}c^\dagger_{\alpha_1\sigma_2}  c_{\alpha_2\sigma_2}c_{\alpha_1\sigma_1},
\end{aligned}
\end{equation}
where $U$ is the Coulomb repulsion between electrons on the same orbital (i.e., the on-site Hubbard interaction), $U'$ is the inter-orbital Coulomb repulsion and $J$ is the exchange interaction between spins on different orbitals (i.e., the Hund exchange). Similarly, through two-centre integrals, we decompose the long-range interaction into:
\begin{equation}
\begin{aligned}
    \Ham^{\rm (int)}_{ij} &= \frac12\sum_{\alpha\sigma}\sum_{\beta\sigma^\prime} V_{\alpha\beta} n_{i\alpha\sigma}n_{j\beta\sigma^\prime} + \frac12\sum_{\alpha\beta}\sum_{\sigma\sigma^\prime} K_{\alpha\beta} c^\dagger_{j\beta\sigma}c^\dagger_{i\alpha\sigma^\prime} c_{j\beta\sigma^\prime}c_{i\alpha\sigma}\\
    &\quad + \frac12\sum_{\alpha\neq\beta}\sum_{\sigma\sigma^\prime} V^\prime_{\alpha\beta} c^\dagger_{i\beta\sigma}c^\dagger_{j\alpha\sigma^\prime} c_{j\beta\sigma^\prime}c_{i\alpha\sigma} + \frac12\sum_{\alpha\neq\beta}\sum_{\sigma\sigma^\prime} K^\prime_{\alpha\beta} c^\dagger_{j\alpha\sigma}c^\dagger_{i\beta\sigma^\prime}  c_{j\beta\sigma^\prime}c_{i\alpha\sigma},
\end{aligned}
\end{equation}
where $V_{\alpha\beta}$ represents the Coulomb interaction and $K_{\alpha\beta}$ is the corresponding exchange interaction; similarly, $V^\prime_{\alpha\beta}$ is the correlation between two on-site exchange interactions, while $K^\prime_{\alpha\beta}$ is the correlation between off-site exchange. 
    
\subsection*{Modeling of the experimental device}
We set the variance of the quantum well potential $\delta = 100$~nm equal to the designed diameter of the quantum dots in the device.$^{4}$ Setting the potential depth $V_0 = 11.4$~meV, we obtain the first-excited-state level spacing $\varepsilon_1-\varepsilon_0 \approx 0.75$~meV. The evaluation of the electron-electron interaction requires a specific value of the dielectric constant, whose bulk value is $\epsilon = 12.9$ in GaAs. However, since the gate electrodes contribute an additional capacitance to the self-capacitance between the dot and the reservoir, we can account for this effect by selecting a larger effective $\epsilon$. Using $\epsilon = 20$ in the quantum-dot system mentioned above, we obtain the ground-state Hubbard interaction $U_0\approx2.34$~meV and the ground-excited-state interaction $U^\prime_{01} \approx 1.92$~meV. This makes the model consistent with the experimental measurements. This multi-orbital ab initio model correctly captures the energy level mixture caused by having the on-site interaction being much larger than the orbital energy-level spacing, a feature that is characteristic of quantum dots.
    
We calculate the long-range interactions for a distance $d=210$~nm between neighbouring dots in the plaquette. The Coulomb interaction $V$ obtained from calculation ranges from 0.22~meV to 0.4~meV depending on the orbitals, $K$ and $V^\prime$ are on the order of or below 1~$\mu$eV, and $K^\prime$ is even lower, on the order of 0.1 or 0.01~$\mu$eV. Even though these higher-order correction terms are much smaller than the on-site interactions, they are still comparable to the $\sim 1$~$\mu$eV high-spin to low-spin energy gap--which we refer to as the Nagaoka gap--and should be taken into account.
    
\begin{figure}
	\includegraphics[scale=0.9]{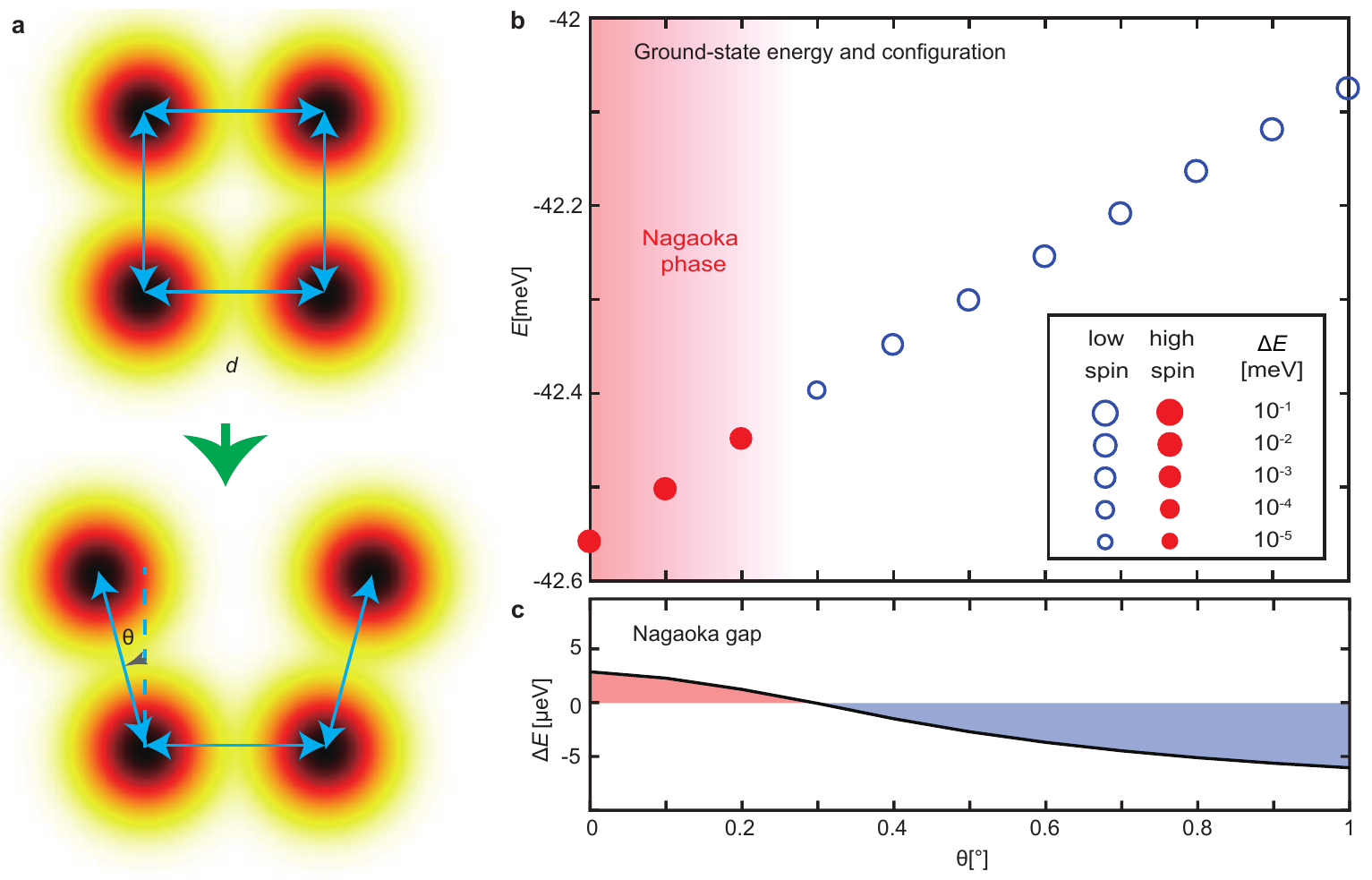}
	\caption{\textbf{Ab initio simulations: 2D to 1D.} \textbf{a,} Schematic of the methodology used in the ab initio simulations to reproduce the effect of the 4-dot system transition from a 2D plaquette to a 1D chain. We gradually vary the angle $\theta$, which effectively varies the distance between two of the dots. \textbf{b,} The ground-state energy and spin configuration, and \textbf{c,} the ferromagnetic to low-spin energy gap $\Delta$ as a function of $\theta$. The ground state soon becomes a low-spin state for the rotating angle at 0.3$^\circ$.}\label{sfig:angle}
	\includegraphics[scale=0.9]{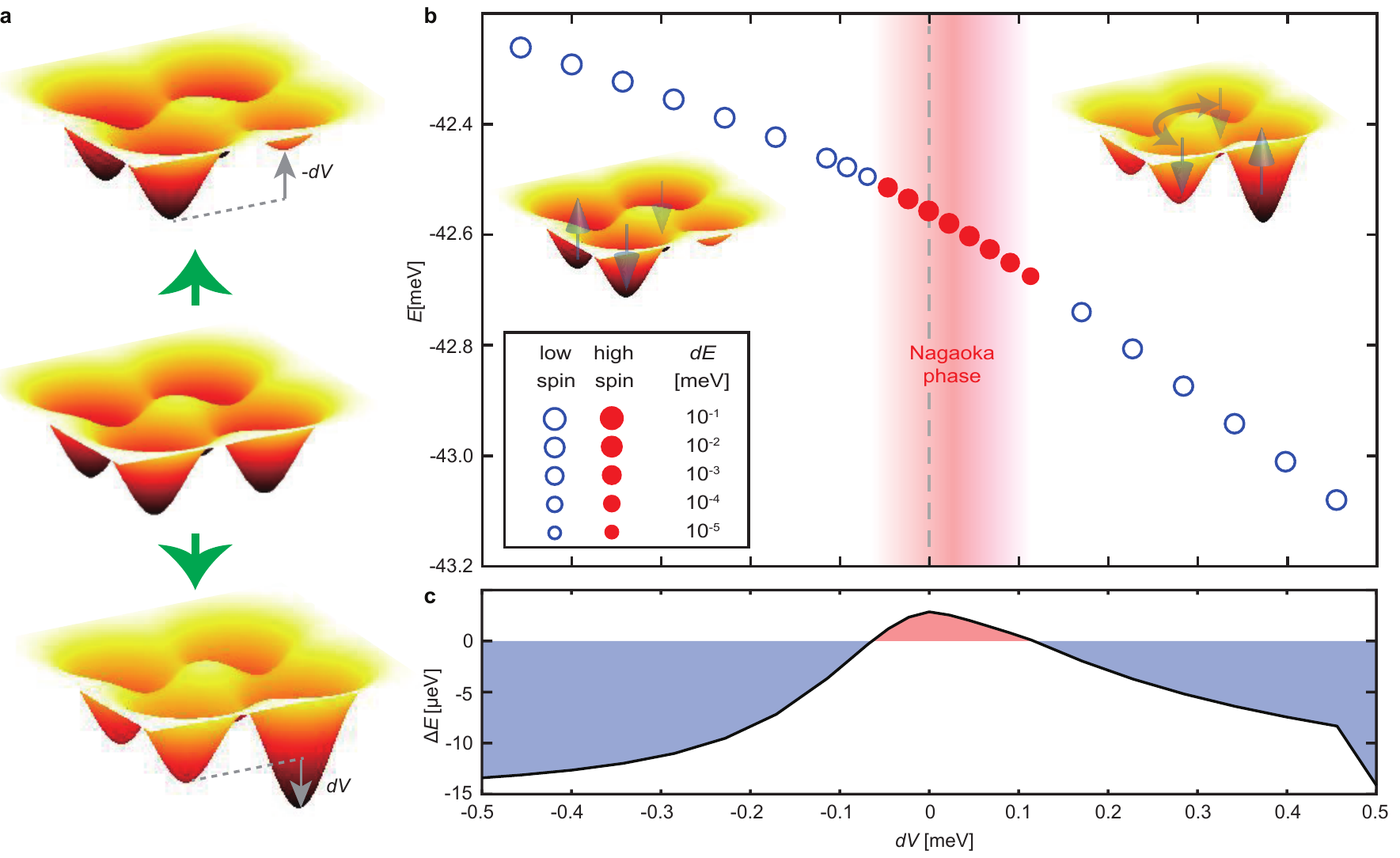}
	\caption{\textbf{Ab initio simulations: local energy offsets.} \textbf{a,} Schematic of the methodology used in the ab initio simulations to reproduce the effect of a local energy offset. The amplitude of the potential of one of the quantum wells is changed by an amount $dV$. The variation of the single-well potential by positive or negative $dV$ gives unbalanced site-energies. Besides, with the change of eigenstate basis, the hybridisation and interaction parameters are also affected in the ab initio calculation. \textbf{b,} The ground-state energy and spin configuration, and \textbf{c,} the ferromagnetic to low-spin energy gap $\Delta$ as a function of $dV$. When the potential detuning is $dV=0.11$~meV or $dV=-0.07$~meV, the system undergoes a transition to a low-spin ground state. The transitions at these two directions have a different nature, as drawn in the insets. For $dV>0$, the particular quantum dot is deeper and tends to trap more electrons. On the other hand, a negative $dV$ raises the energy cost on the particular quantum well and leads to a lower probability of occupation in a three-electron system. Without the ``mobile'' hole in the ``half-filled'' system, the ground state becomes a low-spin state instead a Nagaoka ferromagnetic state.}\label{sfig:detuning}
\end{figure}

For clarity, we distinguish between the \emph{hopping parameter} in the ab initio model, and the experimentally measured \emph{tunnel coupling}. Different from the single-band model described in the previous section, the hopping strength has contributions from all possible paths through different orbitals. The hopping parameters in the tight-binding model vary among different orbitals and typically decrease exponentially as a function of the distance between quantum wells. Since the ground-state wavefunction is most localised, hybridisation between two ground states across neighbouring quantum wells is small ($\sim0.06~\mu$eV for $d=210$~nm). However, with the presence of multiple quantum dots, the tunnel couplings among low-energy states of neighbouring quantum wells--obtained from the superposition of all contributing excited-state paths--become much larger than the bare hopping parameter between ground states. In our ab initio calculation, we estimate the tunnel coupling $t$ by calculating the single-particle bandwidth in the system. Assuming $t$ is dominated by nearest-neighbour tunnelling, the low-energy band structure of a 2$\times$2 plaquette is $-2t\cos \theta$ where $\theta$ goes from 0 to $2\pi$. Therefore, the width of the lowest single-electron band (the lowest four states) is approximately $4t$. For our chosen $d = 210$~nm, the model predicts $t \approx 40$~$\mu$eV, similar to the values measured in the experiment.
    
We perform the ab initio, exact-diagonalisation calculation, with three electrons in a four-well system, emulating the experimental conditions. The bottom-level differential equation and integration are calculated on a grid with a spacing of 1nm. To simplify the calculation, we keep 15 orbitals in each quantum well, which span a $\sim 5$~meV energy range, much larger than both $U$ and $t$. The solution indeed predicts a high-spin ground state, with a Nagaoka gap of $\sim 3$~$\mu$eV.
    
We have reproduced two of the experiments described in the main text. We first model the transition of the 4-dot array from a ring to a chain, by gradually increasing the distance between two of the dots (Fig.~\ref{sfig:angle}). This effectively reduces the tunnelling term between them, with a transition to a low-spin ground state when the system becomes more 1D-like, as described in the main text.
    
Finally, we reproduced the effect of varying the local energy offset, by gradually varying the amplitude of the potential of the quantum well in one of the dots (Fig.~\ref{sfig:detuning}). The model predicts transitions to a low-spin ground state for both positive and negative local offsets. As observed in the experiment, these transitions occur over a range of energy orders of magnitude larger than the tunnel coupling, and with a similar asymmetry between positive and negative offsets.

\section*{Local energy offsets: additional figures}
\begin{figure}[b]
	\includegraphics[scale=0.85]{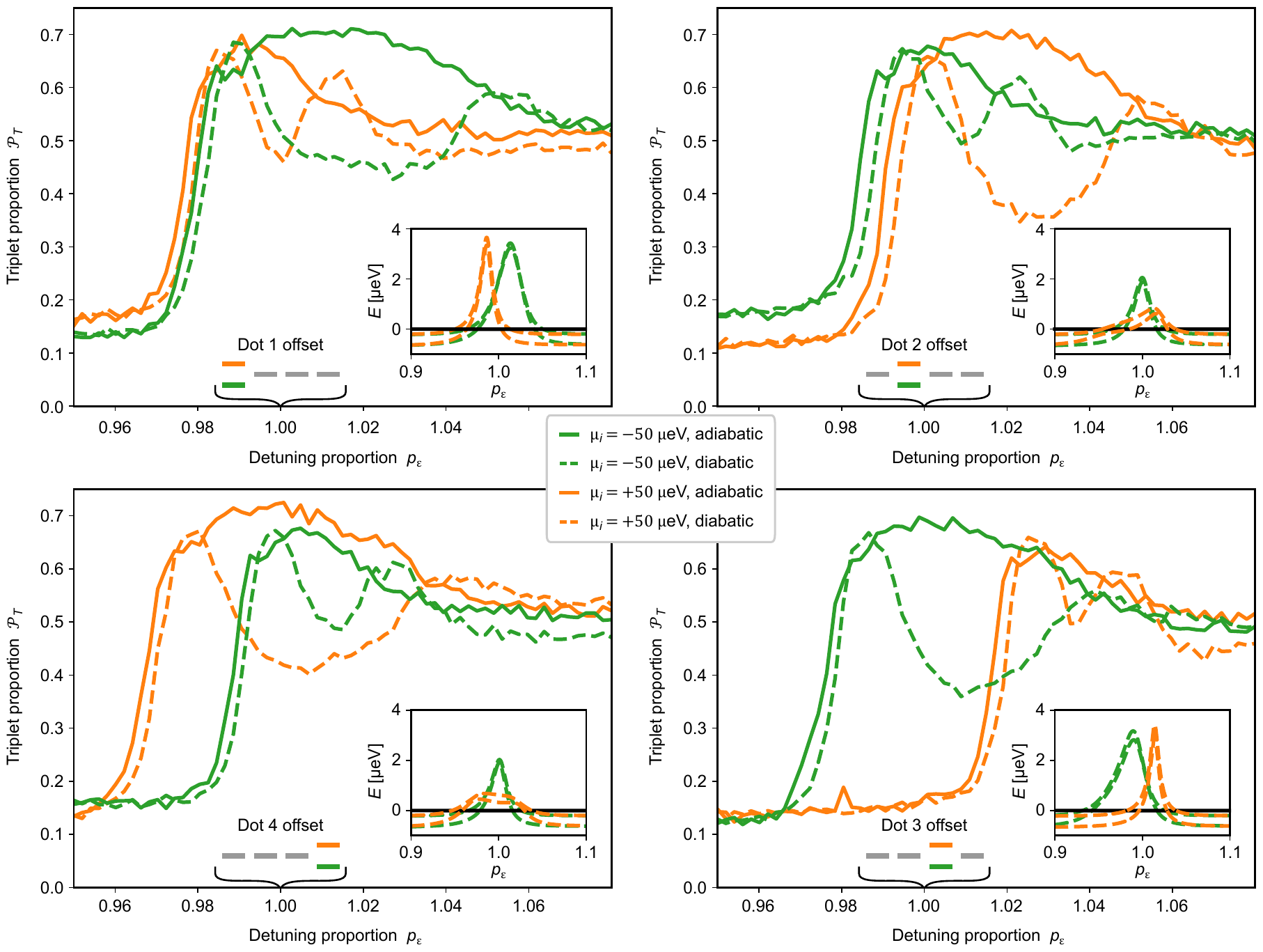}
	\caption{\textbf{Local energy offsets on all dots.} Same measurement as in Fig.~6, applying the $\pm 50$~$\mu$eV offset on each of the four dots. Panels correspond to offsets in dots 1 to 4, clockwise from the top-left. Note that the asymmetry in the plots is related to the fact that the local energies at point M are in an asymmetric detuning configuration and we pulse linearly from this configuration to point N. As expected, the simulated energies of the different spin states at point N ($p_\epsilon = 1$), are the same in all four plots.}\label{sfig:4offsets}
\end{figure}
\clearpage
\begin{figure}
	\includegraphics[width=\textwidth]{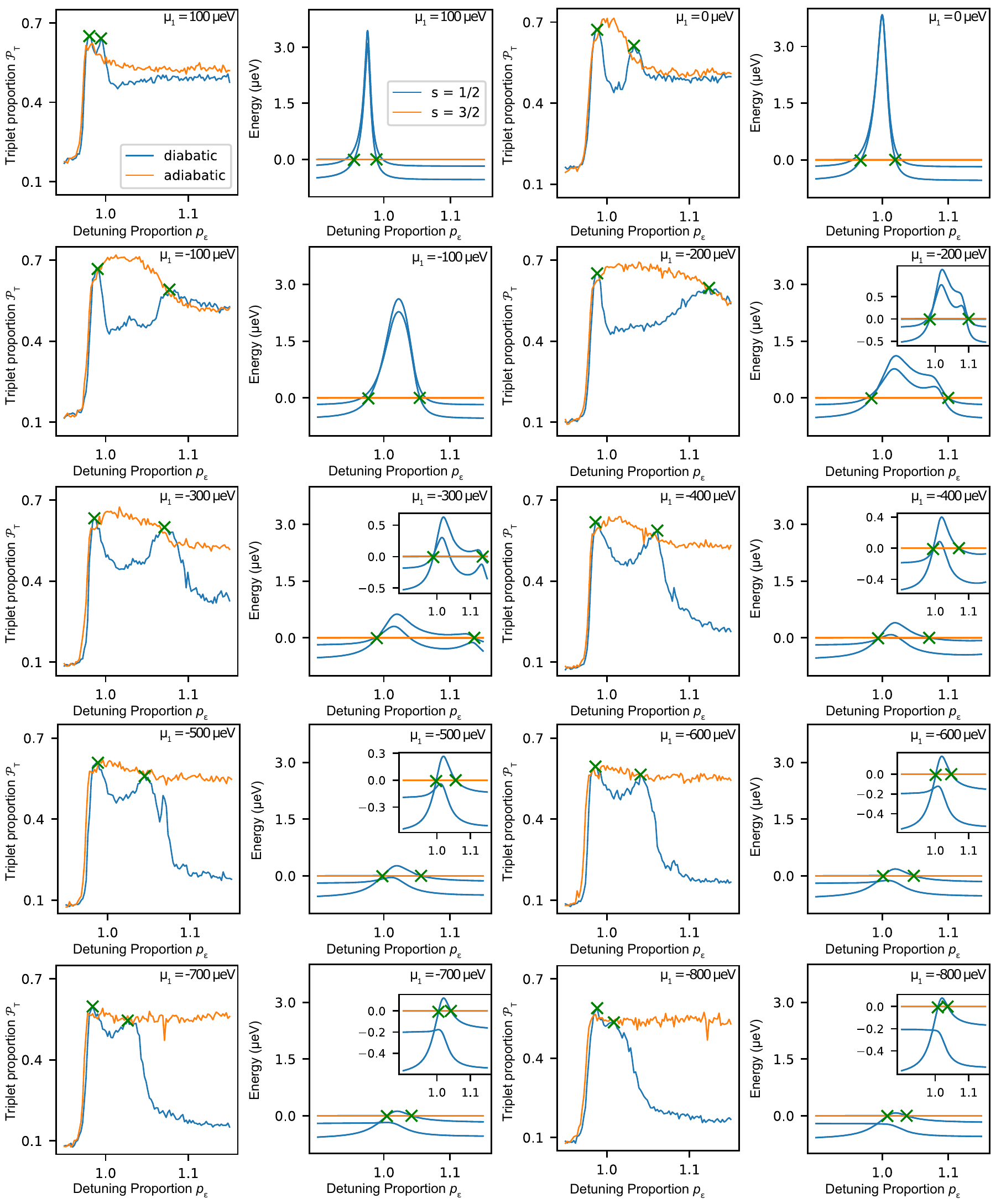}
	\caption{\textbf{Large local offsets.} Each pair of panels show experimental measurements (left) and simulated spectra (right), where point~$N$ has been redefined such that the chemical potential of dot~1 is offset by the amount shown on the top right of each panel. Green \textit{X}s highlight the detuning points used to obtain the values in Fig.~6b. For experimental plots, these points where obtained using a peak-finding algorithm (local maxima by simple comparison with neighbouring values); for simulated plots, the points correspond to the energy level crossings.}\label{sfig:largeoffsets}
\end{figure}
\clearpage

